\documentclass[11pt]{article}

\usepackage{amsmath}
\usepackage{amsfonts}
\usepackage{amssymb}
\usepackage{graphicx}
\usepackage{epstopdf}
\usepackage{cite}
\usepackage[usenames]{color}

\providecommand{\U}[1]{\protect\rule{.1in}{.1in}}
\newcommand{\ba}{\begin{array}}
\newcommand{\ea}{\end{array}}
\newcommand{\Dsl}[1] { \setbox0=\hbox{$#1$}     
\dimen0=\wd0   \setbox1=\hbox{/} \dimen1=\wd1  \ifdim\dimen0>\dimen1        
 \rlap{\hbox to \dimen0{\hfil/\hfil}}  #1 \else \rlap{\hbox to \dimen1{\hfil$#1$\hfil}}  /  \fi  }
\newcommand{\bea}{\begin{eqnarray}}
\newcommand{\eea}{\end{eqnarray}}

\newcommand{\ns}{\Dsl{n}}
\newcommand {\nbs}{\Dsl{\bar n}}

\newcommand{\lab}{\label}

\begin{document}
\title{ \Large \bf Two photon exchange corrections at large momentum transfer revised\footnote{Submitted to Acta Physica Polonica B special volume dedicated to
Dmitry Diakonov, Victor Petrov and Maxim Polyakov} }
\author{ Nikolay Kivel
\\ %[1cm]
\textit{ Physik-Department, Technische Universit\"at M\"unchen,}\\ \textit{James-Franck-Str. 1, 85748 Garching, Germany } 
 }

\maketitle

\begin{abstract}
Motivated by experimental data at large momentum transfer 
we update the analysis of the two-photon exchange effect  in the  electron-nucleon scattering using the effective field theory formalism.  Our approach is suitable for describing the hard $s\sim |t|\sim |u|\gg \Lambda^2$ region, where the hadronic model calculations are not accurate enough. We improve the estimates of various long-range matrix elements and discuss the obtained numerical effects for the unpolarised elastic cross section.

Assuming a linear behaviour of the reduced cross section with respect to the photon polarisation, we show that the obtained  description allows us  to resolve the  form factor discrepancy  
for  $Q^2=2.5–3.5\,$GeV$^2$ . However, the effect obtained is quite small for higher values of $Q^2$. It is possible that nonlinear effects may be important in understanding the discrepancy in this region.

Estimates of the elastic electron-neutron cross section in the region $Q^2=2.5-3.5\,$GeV$^2$ are also performed. The obtained TPE effects are  sufficiently large and must be taken into account.    
\end{abstract}
\noindent

\vspace*{1cm}
\newpage

\section*{Dedicated to Mitya, Vitya and Maxim}
I first met Maxim Polyakov back at Saint Petersburg University, apparently in 1989. The Faculty of Physics and the Institute of Physics were located in Peterhof, a suburb of Saint Petersburg, and we both lived on campus there, in the student dormitories.

Maxim was already a senior student and participated as an equal in scientific discussions with the department staff. I was just starting my fifth semester, so I could only attend as a listener, as I didn’t yet understand anything in those discussions. Terms like “pions,” “K-mesons,” and “SU(3) violation” were completely unclear to me. This strongly motivated me to pursue self-education, developing independently of the basic curriculum. At that time, Maxim was writing his diplom thesis at the Petersburg Institute of Nuclear Physics under the supervision of two individuals unknown to me then, Mitya and Vitya. This was how I first became indirectly acquainted with D.I. Dyakonov (Mitya) and V.Yu. Petrov (Vitya).

Later, I joined the group of Prof. A.N. Vasiliev, who worked on quantum field theory and its applications in statistical physics. After defending my diploma at the end of 1992, I entered the PhD program in the theoretical department of PNPI. At that time, I was  eager to switch to interesting problems in elementary particle physics.

One of the opportunities to better understand the research topics of the theoretical department and get to know its members was the in-house theoretical seminar. It was during these seminars that I first met Mitya and Vitya in person. These seminars were not the typical 45-minute sessions common almost everywhere today. They were held every Monday and could last up to four hours with a short break. This seminar was a characteristic feature of the theoretical department, and it seems the tradition originated with V. Gribov, who inherited it from the seminars of Academician Lev Landau. During the speaker’s presentation, it was normal to constantly ask questions, critique, demand additional explanations, disagree with certain points by explaining the reasoning to everyone, or, conversely, provide alternative explanations and diverse commentary. Often, complex debates arose among participants that sometimes became unclear even to the speaker. Overall, it was a lively forum — a place for discussion where the speaker faced extensive criticism, sometimes constructive, sometimes not, and had to clearly explain and defend his methods and conclusions while considering various perspectives and approaches from the audience.

Undoubtedly, the most active participants, including Vitya and Mitya, formed the core of this forum. Speakers often received the most questions and criticism from Vitya, who sometimes reacted passionately to claims or hypotheses he deemed incorrect. “But Landau and Lifshitz already wrote about this in third volume (Quantum Mechanics)!” he would often exclaim. Mitya was not far behind, though he spoke more calmly, usually presenting very strong arguments. To convince them, a speaker not only needed a deep understanding of the topic but also had to be able to clearly explain where Mitya and Vitya might be mistaken.

It took me considerable time to properly grasp the terminology and the commonly used working jargon, and even after that, I still felt uncertain for a long time. Somehow, it so happened that I found myself without a scientific advisor for an extended period. By that time, many  researchers had already left to work in various laboratories abroad, and finding a topic that would interest and captivate me turned out to be difficult. Thus, I continued working on my diploma thesis topic with the university group, focusing more on mathematical issues in statistical models of quantum field theory.

In the fall of 1993, after one of the seminars, Mitya approached me:
“We have a tradition,” he said, “we assign young PhD students an overview talk on a completely new and unfamiliar topic they have not worked on before. So we’ve decided to ask you to prepare a talk on the heavy quark effective theory (HQET). This topic is very popular now, and it will be very useful for everyone to learn more about  it.”

I must admit, this hit the mark precisely — I knew absolutely nothing about the physics and phenomenology of heavy quarks! Moreover, at that time I had not yet published anything at all on elementary particle physics. For this reason, the complexity of the task sounded to me almost like the famous phrase from a Russian fairy tale: “Go there, I don’t know where, bring that, I don’t know what.” I felt uneasy.
“Could you recommend anything for reading and preparation?” I asked Mitya.
“Talk to Kolya Uraltsev; he’s just back for a short while and will be in the theoretical department on Thursday.”

That didn’t make me feel much better. I had only heard of Kolya but had not met him personally, as he was almost always away working abroad. I managed to meet Kolya at the institute only once. He was very busy with some administrative matters, quickly handed me two preprints without much explanation, and rushed off. That was the entirety of our interaction. The preprints turned out to be original articles on B-meson decays, intended for an advanced reader, and I found almost nothing useful in them for myself. Even the first formula for the effective heavy quark field was unclear to me — its derivation and reasoning were not provided in detail.

Tracing references in those papers also yielded no results, as they were similar original works intended for advanced readers. Ultimately, it seemed I would need to independently derive nearly all the fundamentals of the effective theory, which was, of course, unrealistic. I became despondent, as I could not find a reasonable way to prepare for the talk.

Another psychological factor added to the pressure. Mitya was known for his plain dealing in evaluations. He sincerely believed that if someone, in his opinion, is not suitable for research work, then it is better to tell him about it directly.
Usually this sounded like a recommendation not to engage in science. Given Mitya’s authority, such a recommendation was not easy to hear. I personally witnessed one such conversation with a student who had incorrectly solved a problem during the theoretical minimum exam to enter the PhD program in the theoretical department. Since Mitya had personally assigned me the task, I feared that if I did not prepare adequately, I would face a similar fate.

Fortunately, the situation with the talk was eventually resolved. I can’t remember how, but I found a reference to a recently published review by M. Neubert titled "Heavy Quark Symmetry", available in the electronic hep archive \cite{Neubert:1993mb}. It was exactly what I needed — a superb introduction to the physics of heavy quarks and specifically HQET, which I needed to present to the seminar audience. With this discovery, my stressful struggle ended, and the preparation became genuinely enjoyable. I delved deeply into the physics and technical details, reproduced some calculations myself, and successfully prepared and delivered the seminar. This was essentially my first experience of independently mastering a completely new and unfamiliar topic, which gave me confidence and provided knowledge that proved valuable later.

Several years later, while in Germany, I more than once met with Mitya and Vitya in Bochum, where I was working at the time in Prof. Klaus Goeke’s group. They were actively collaborating with Klaus and Maxim on various topics, such as pentaquark and various applications of the quark-soliton model. One day, I mentioned to Vitya that there was still a lack of data on hard exclusive processes. To my surprise, he disagreed, arguing that, in his view, the available data were sufficient, but that the major gap is in theory — we don’t understand confinement and can’t systematically calculate all necessary quantities in QCD from first principles. He and Mitya were actively working on the understanding of the confinement in QCD, and Vitya firmly believed that it was purely a theoretical challenge. To me, it seemed strange that such a fundamental issue could exist independently of experiments, but on the other hand, I couldn’t propose a way to experimentally study confinement either. However,  I believed that our lack of understanding of low-energy dynamics might be due to insufficient data that could guide us to the right solution, so experimental exploration is also crucial. From this point of view, hard exclusive  processes are especially interesting because they involve  a combination of QCD dynamics at small and large distances and provide access to a new information.  We started a discussion, but of course I couldn’t convince Vitya.

Despite this somewhat skeptical point of view, Vitya in collaboration  with Maxim produced a number of interesting works on calculating hadron wave functions (or, more precisely, light-cone distribution amplitudes, LCDAs) in the quark-soliton model. These functions are very important for description various hard exclusive reactions.  One of their papers focused on nucleon LCDAs and the calculation of the proton’s electromagnetic form factors. For reasons unknown to me, these results were only published as an electronic preprint. Several years later, I needed different models for nucleon LCDAs for my work, but I couldn’t find the parameterisation I required in their paper. I approached Maxim, explaining that there were significantly different estimates of an important constant obtained through QCD sum rules and lattice calculations. Therefore, it would be very interesting to see what value this constant could have in the quark-soliton model.Maxim promised to calculate the constant, and just a few days before his tragic passing, he told me that the preliminary result was ready, needing only verification and evolution to the relevant scale for comparison with other results. This work was later completed and published by his colleagues.

In terms of scientific collaboration, my closest interaction is with Maxim.
Most of our joint works, with rare exceptions, were related to the study of Generalised Parton Distributions (GPDs) and understanding of Deeply Virtual Compton Scattering. While I was primarily interested in process factorisation and the calculation of radiative and power corrections, Maxim was more focused on the properties of non-perturbative matrix elements. In this way, we complemented and learned from each other. It was thanks to him that I became interested in chiral perturbation theory and its application to the description of non-perturbative functions in hard processes. One of the first ideas of this kind was the study of pion GPDs and soft-pion low-energy theorems for these functions. Specifically, the chiral calculation automatically provided the correct structure of the double distribution for GPDs, including the so-called D-term, which, for example, is absent in the naive model associated with the triangle diagram.

When we started this project, I was completely unfamiliar with chiral perturbation theory, so I had to draw on my prior experience studying HQET, as described earlier. I found several reviews, studied them thoroughly, and quickly joined the work. This experience enriched my understanding of the effective field theory approach and proved valuable for other applications later.

Further study of chiral loop corrections led us to conclude that the chiral expansion for GPDs breaks down in the region of small Bjorken $x_{Bj}$. The resulting problem, in principle, could only be resolved by resumming all orders of the chiral expansion. At first glance, this task seemed insurmountable, but we quickly realised it could be significantly simplified by assuming that the dominant contribution came from leading chiral logarithms. In this case, it turned out that the pion could be treated as massless, and we could focus only on four-particle vertices with derivatives in the full chiral Lagrangian. Essentially, the problem reduced to one-loop renormalisation of an infinite number of effective constants in this effective action. However, even in this case, renormalisation of the vertices remained complex and confusing due to their strong mixing with one another.
Here, the experience I gained while studying critical statics proved helpful. The idea emerged to introduce a conformal basis for vertex operators, which greatly simplified the mixing structure, allowing us to derive a relatively simple recurrence relation for the coefficients of the large leading logarithms. I must say that working on this problem gave me immense satisfaction from solving mathematical challenges. It was especially paradoxical for me to see the problem reduced to conformal field theory. After all, at first glance, it was difficult to imagine that there could be any connection between conformal symmetry and chiral perturbation theory! Maxim was also fascinated, primarily by the mathematical aspects of our approach. Unfortunately, it turned out that from a practical standpoint, the effect of logarithm resummation was relatively small, though it provided some qualitative insight into the behaviour of the chiral pion cloud in the small $x_{Bj}$ region.
Later, I became interested in calculating the two-photon contribution to elastic scattering and paused further work on this project. However, the experience gained in calculating chiral corrections and, more broadly, the perspective on the concept of low-energy effective theories became an integral part of my practical toolkit and proved extremely useful for solving some problems later, for which I am sincerely grateful to Maxim.

\newpage
\section{Introduction}

The two-photon exchange (TPE) contribution gives us the most plausible  explanation of the discrepancy
 in extraction of the  proton  ratio $G_{E}/G_{M}$  using the Rosenbluth (or LT separation ) and the
polarisation transfer methods, see {\it e.g.} reviews \cite{Perdrisat:2006hj, Carlson:2007sp,Arrington:2011dn,Afanasev:2017gsk,Borisyuk:2019gym}. 
Many different TPE calculations have been
carried out  in  the region of $Q^2<5\ $GeV$^{2}$.
This includes the various hadronic model calculations
 \cite{Blunden:2005ew,Kondratyuk:2005kk,Kondratyuk:2007hc, Zhou:2014xka} and also
dispersive framework \cite{Borisyuk:2008es,Borisyuk:2012he,Borisyuk:2015xma,Tomalak:2014sva,Blunden:2017nby,Ahmed:2020uso}. 
These estimates  allow one to obtain the
TPE contribution at relatively low $Q^2\leq 2 $GeV$^2$. 
However, such calculations become less accurate in the region of higher values of the momentum transfer, since they do not properly take into account the effects of short-range dynamics.
This become especially important in the  region where all Mandelstam variables describing the $ep$-scattering  are large
\bea 
s\sim -t\sim -u\gg \Lambda^2,
\label{def:hard}
\eea
 where  $\Lambda$ is the typical hadronic scale. For any  large $Q^2$ there is the region  $\varepsilon>\varepsilon_{min}$  where  condition (\ref{def:hard}) is satisfied.  In this region the calculations of the hadron model have a large uncertainty from the contribution of excited states. With the growth of $Q^2$, the value of $\varepsilon_{min}$ decreases, limiting more and more the applicability domain of the hadron model.

Polarisation measurements of $\mu_{p}G_{E}/G_{M}$ show that this ratio decreases monotonically with increasing $Q^{2}$ and becomes quite small in the region of $Q^{2}\sim 9-10$ GeV$^{2}$.  At the same time the LT separation approach suggests  that this ratio remains large $\mu_{p}G_{E}/G_{M}  \sim1-2$ \cite{Arrington:2003qk, Christy:2021snt}. This discrepancy could be interpreted as the TPE correction being large and dominant in this kinematic region.  A  description of the TPE amplitudes  in this case must  include  the  effects from the short  and long distance dynamics in a systematic way.    

The collinear factorisation of TPE amplitudes  at large momentum transfer were considered  in Refs. \cite{Borisyuk:2008db, Kivel:2009eg}.  Such a  description includes
only the hard-spectator mechanism. The experience in the description of proton FFs  allows on to expect that the
soft-spectator contributions are also large \cite{Isgur:1984jm, Braun:2006hz, Kivel:2010ns} and therefore they can also
provide sufficiently large numerical effect. The  first attempt to consider such mechanism was done in Ref.\cite{Afanasev:2005mp} using the GPD-model framework.  A more systematic treatment  based on the
effective field theory (EFT) framework was considered in Ref.\cite{Kivel:2012vs}. 

n the EFT approach  the soft-spectator contribution is described in
terms of  appropriate soft-collinear matrix elements  associated with the soft and  collinear particles. 
Corresponding  momenta  are of  order $p_{s}^{2}\sim\Lambda^{2}$ and $p_{hc}^{2}\sim\Lambda Q$ for the soft and hard-collinear particles, respectively. These contributions are considered as non-perturbative.  The hard interactions associated with the momenta  $p_{h}^{2}\sim Q^{2}$ are computed  using the perturbative QCD (pQCD). Such an approach allows one to include  hard- and soft-spectator contributions on a systematic way. 

In the current work we update the numerical calculations carried out in
Ref. \cite{Kivel:2012vs} for the reduced cross section.  We are also going to clarify some uncertainties
associated with  long distance  matrix elements. The TPE effect in the cross section with  a neutron target is also discussed. 

The paper is organised as follows.  In Sec.$\,$\ref{description}  we discuss the analytical structure of the TPE amplitudes,  briefly review the results obtained previously for these amplitudes and provide 
an explicit formula for the unpolarised cross section used in  the phenomenological analysis.  In Sec.$\,$\ref{proton}
we carry out the phenomenological analysis of the existing data  for the elastic electron-proton scattering. 
The Sec.$\,$\ref{neutron} is devoted to the TPE  calculations  for the   elastic electron-neutron scattering.
In Sec.$\,$\ref{disc} we briefly discuss the obtained results and conclusions.
In the three Appendices we provide some important technical details.

In Section $\,$\ref{description} we discuss the analytical structure of the TPE amplitudes, briefly review the results obtained previously for these amplitudes, discuss their properties, and provide an explicit formula for the unpolarized cross section used in the phenomenological analysis. In Section $\,$\ref{proton} we perform a phenomenological analysis of the existing data for elastic electron-proton scattering. Section $\,$\ref{neutron} is devoted to TPE calculations for elastic electron-neutron scattering. In Section $\,$\ref{disc} we briefly discuss the obtained results and conclusions. In three appendices we provide some important technical details.
\bigskip

\section{The TPE contribution within the EFT framework}
\label{description}
In this paper we use the same notation as in Ref.\cite{Kivel:2012vs}. For simplicity, we will consider the description for a proton target. Specific features for a neutron target will be discussed later.

The expression for the TPE amplitude is given by
\begin{equation}
A_{ep}^{\gamma\gamma}=\frac{e^{2}}{Q^{2}}\bar{u}(k^{\prime})\gamma_{\mu
}u(k)~\bar{N}(p^{\prime})~\hat{T}^{\mu}(\varepsilon,Q^{2})~N(p),
\label{def:Agg}%
\end{equation}
where $\bar{u}(k^{\prime}),u(k)$ and $\bar{N}(p^{\prime}),N(p)$ denote 
lepton and nucleon spinors, respectively.  The amplitude $\hat{T}^{\mu}$ reads   \cite{Guichon:2003qm}
\begin{equation}
\hat{T}^{\mu}(\varepsilon,Q^{2})=\gamma^{\mu}\delta\tilde{G}_{M}^{2\gamma
}(\varepsilon,Q^{2})-\frac{P^{\mu}}{m}\delta\tilde{F}_{2}(\varepsilon
,Q^{2})+\frac{P^{\mu}}{m^{2}}\Dsl{K} ~\tilde{F}_{3}(\varepsilon,Q^{2}),
\end{equation}
where
\begin{equation}
P=\frac{1}{2}(p+p^{\prime}),~\ K=\frac{1}{2}(k+k^{\prime}),
\end{equation}
and $m$ denotes the nucleon mass. 
Variables $\varepsilon$ and $Q^{2}$ denotes the photon polarisation and
momentum transfer as usually. The reduced elastic cross section is given by
\begin{equation}
\sigma_{R}^{el}(\varepsilon,Q)=G_{M}^{2}+\frac{\varepsilon}{\tau}G_{E}
^{2}+2G_{M}\operatorname{Re}\left[  \delta\tilde{G}_{M}^{2\gamma}
+\varepsilon\frac{\nu}{m^{2}}\tilde{F}_{3}\right]  +2\frac{\varepsilon}{\tau
}G_{E}\operatorname{Re}\left[  \delta\tilde{G}_{E}+\frac{\nu}{m^{2}}\tilde
{F}_{3}\right]  . \label{sigamR1}
\end{equation}
where~$\tau=Q^{2}/4m$, $\nu=(K\cdot P)~$, $G_{M}$ and $G_{E}$ are
magnetic and electric form factors (FFs), respectively.  Following the arguments discussed in Ref.\cite{Kivel:2012vs}, we neglect the second term on the {\it rhs} of equation (\ref{sigamR1}), assuming that the numerical effect of this term is sufficiently small due to the small $G_{E}$
\begin{equation}
G_{M}\operatorname{Re}\left[  \delta\tilde{G}_{M}^{2\gamma}+\varepsilon
\frac{\nu}{m^{2}}\tilde{F}_{3}\right]  \gg \frac{\varepsilon}{\tau}%
G_{E}\operatorname{Re}\left[  \delta\tilde{G}_{E}+\frac{\nu}{m^{2}}\tilde
{F}_{3}\right]  \sim O(\alpha G_{E}). \label{GEsmall}
\end{equation}
Therefore, with very good accuracy, the reduced cross-section is determined by the expression
\begin{equation}
\sigma_{R}^{el}(\varepsilon,Q)\simeq G_{M}^{2}(Q^{2})+\frac{\varepsilon}{\tau
}G_{E}^{2}(Q^{2})+2G_{M}(Q^{2})\operatorname{Re}\left[  \delta\tilde{G}
_{M}^{2\gamma}(\varepsilon,Q)+\varepsilon\frac{\nu}{m^{2}}\tilde{F}
_{3}(\varepsilon,Q)\right] .
 \label{sigmaR}
\end{equation}
The interference term in the {\it right-hand side} of the equation (\ref{sigmaR}) is proportional to the small electromagnetic coupling $\alpha$, but for large values of $Q^2$ (large $\tau$) 
this contribution can have a strong influence on the extraction of the small FF $G_{E}^{2}$.

 In the forward limit, when $\varepsilon\rightarrow1$ and $Q^{2}$ is fixed\footnote{This is equivalent in Mandelstam variables to the limit $s\to \infty$ and $-t$ is fixed.}  the  TPE correction vanishes
\begin{equation}
\left.  \operatorname{Re}\left[  \delta\tilde{G}_{M}^{2\gamma}(\varepsilon
,Q)+\varepsilon\frac{\nu}{m^{2}}\tilde{F}_{3}(\varepsilon,Q)\right]
\right\vert _{\varepsilon\rightarrow1}=0, \label{fwd}%
\end{equation}
as it follows from the dispersion relations derived in Ref.\cite{Borisyuk:2008es}.

In the kinematic region, where all Mandelstam variables are large $s\sim-t\sim-u\gg\Lambda^{2}$, the complex dynamics of QCD is sensitive to different scales associated with different scattering mechanisms.  The amplitudes are given by the sum of the soft- and hard-spectator scattering contributions, which are designated  by $(s)$ and $(h)$ superscripts, respectively 
\begin{equation}
\delta\tilde{G}_{M}^{2\gamma}(\varepsilon,Q)=\delta\tilde{G}_{M}%
^{(s)}(\varepsilon,Q)+\delta\tilde{G}_{M}^{(h)}(\varepsilon,Q),
\end{equation}%
\begin{equation}
\tilde{F}_{3}(\varepsilon,Q)=\tilde{F}_{3}^{(s)}(\varepsilon,Q)+\tilde{F}%
_{3}^{(h)}(\varepsilon,Q).
\end{equation}
The  hard-spectator contributions can be calculated  using  the QCD collinear factorisation approach \cite{Borisyuk:2008db,Kivel:2009eg}.  The analytical results  are collected  in Appendix~\ref{appA}.

\begin{figure}[ptb]
\centering
\includegraphics[width=2.0in]{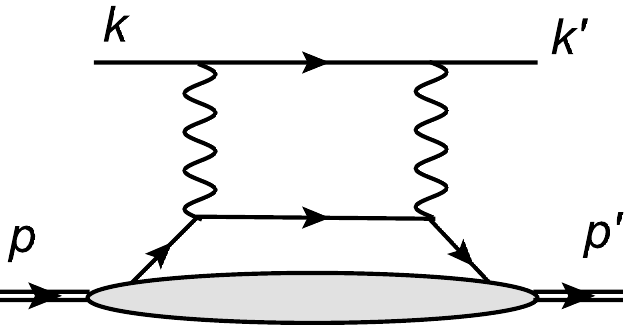}
\caption{ The soft-spectator contribution.  The grey blob denotes  FF $\mathcal{F}_1(Q^2)$ }
\label{soft-spectator}
\end{figure}

The soft-spectator contributions  are described by the product of the hard coefficient functions and long distance matrix elements. The hard subprocess is associated with  lepton-quark subprocess $eq\rightarrow eq$. In this
case the both photons couple to a single active quark giving the box diagrams as in
Fig.\ref{soft-spectator}.  It is assumed that the virtual photons and quark in this diagram have large virtualities
of order $Q^{2}$. 

 Therefore the soft-spectator contribution $\tilde{F}_{3}^{(s)}$ can be
represented as
\begin{equation}
\frac{\nu}{m^{2}}\tilde{F}_{3}^{(s)}(\varepsilon,Q^{2})=\frac{\alpha}{\pi}\,
\frac{\nu}{s}C_{3}(\varepsilon,Q^{2})\,\mathcal{F}_{1}(Q^{2}),
\label{F3}%
\end{equation}
where $C_{3}(\varepsilon,Q^{2})$ is the hard coefficient function describing
the hard lepton quark subprocess $eq\rightarrow eq$.  The non-pertubative
FF $\mathcal{F}_{1}(Q^{2})$ describes the hard-collinear matrix element and 
 includes the interactions of the soft and the hard-collinear modes in
the EFT. This function can be restricted from the data for wide angle Compton
scattering (WACS)  \cite{Kivel:2012vs,Kivel:2015vwa}. 

The expression for  $\delta\tilde{G}_{M}^{(s)}$ includes  two different
contributions. One term describes a configuration in which both the photons and the loop quark are hard, as noted earlier.  The second term  is associated with the region, where  photons  have 
different virtualities: hard $q_{1}^{2}\sim Q^{2}$ and soft $q_{2}^{2}
\sim\Lambda^{2}$ or vice versa. Therefore  the factorised expression reads%
\begin{equation}
\delta\tilde{G}_{M}^{(s)}(\varepsilon,Q^{2})=\frac{\alpha}{\pi}C_{M}%
(\varepsilon,Q^2,\mu_{F})\mathcal{F}_{1}(Q^{2})+\frac{\alpha}{\pi}%
g_{1}(\varepsilon,Q^2,\mu_{F}).~\label{dGM}%
\end{equation}
Here the hard loop contribution  is associated with the  $C_{M}~\mathcal{F}_{1}$ term and  $C_{M}$ is the hard coefficient function, which can be found in Appendix~\ref{appB}.

The amplitude $g_{1}$ describes the contributions with the soft photon. The hard-hard and hard-soft photon configurations overlaps and  the factorisation scale $\mu_{F}$ separates corresponding domains.  It can be associated with maximal virtuality of the soft photon.

 The long distance amplitude $g_{1}$ is defined  by the matrix element in the EFT after factorisation of the hard subprocess.  It is important to notice  that  this amplitude 
includes the infrared QED logarithm, which arises from the TPE diagrams. In
Ref.\cite{Kivel:2012vs} the elastic contribution to  $g_{1}$ is computed using  the effective hadronic model. The result reads
\begin{equation}
g_{1}(\varepsilon,Q,\mu_{F})\simeq G_{M}(Q^{2})\ln\frac{\lambda^{2}}{\mu
_{F}^{2}}\ln\left\vert \frac{s-m^{2}}{u-m^{2}}\right\vert +g_{1R}(\varepsilon,Q),\label{g1el}%
\end{equation}
 where $\lambda^{2}$ is the photon mass used as
a IR-regulator, $s$ and $u$ are the Mandelstam variables, which can be
rewritten in terms of $\varepsilon$ and $Q^{2}$  (recall, $m$ is nucleon mass). The function $g_{1R}$ in
Eq.(\ref{g1el}) includes inelastic IR-finite contributions. Such contributions
are not considered in Ref. \cite{Kivel:2012vs}.  It turns out that at large $Q^{2}$ such
contributions are  suppressed and therefore 
\begin{equation}
g_{1R}(\varepsilon,Q,\mu_{F})\approx0.
\end{equation}
A more detailed discussion of  this point is provided in  Appendix~\ref{appB}.  Therefore
the elastic approximation in Eq.(\ref{g1el}) practically gives the dominant
numerical effect in this case. 

In order to compare our calculations with the experimental data, which have been
corrected according to  Mo and Tsai (MT) \cite{Mo:1968cg},   we need to consider the following  expression \cite{Kivel:2012vs}
\bea
\sigma_{R}^{1\gamma,\text{MT}}=\sigma_{R}^{1\gamma}\left(  1+\delta_{2\gamma
}-\delta_{2\gamma}^{\text{MT}}\right)  ,
\label{sgmRfin}
\eea
where $\sigma_{R}^{1\gamma,\text{MT}}$ is the reduced cross section obtained
using MT radiative corrections, $\delta_{2\gamma}$ denotes the TPE correction
obtained in EFT framework,  $\delta_{2\gamma}^{\text{MT}}$ is the expression
obtained in Ref.\cite{Mo:1968cg}. The analytical expression for the factor $\delta
_{2\gamma}^{\text{MT}}$ reads
\begin{align}
\operatorname{Re}\delta_{2\gamma}^{\text{MT}} (\varepsilon,Q^2)&  =\frac{2\alpha}{\pi}\ln\frac{\lambda^{2}}{\tilde{s}}\ln\left\vert
\frac{\tilde{s}}{\tilde{u}}\right\vert +\bar{\delta}_{2\gamma}^{\text{MT}}(\varepsilon,Q^2) ,\label{dltMT}% 
\end{align}
with
\begin{align}
\bar{\delta}_{2\gamma}^{\text{MT}}&  =\frac{2\alpha}{\pi}\left\{
\frac{1}{2}\ln^{2}\left\vert \frac{\tilde{s}}{\tilde{u}}\right\vert 
-\frac{1}{2}\ln^{2}\frac{\tilde{s}}{s}
-\text{Li}\left(  \frac{\tilde{s}}{s}\right)  
-\text{Li}\left(  \frac{u}{\tilde{u}}\right)+\frac{\pi^2}3
\right\}.
\label{dltMTs}
\end{align}
where $\lambda^{2}$ is the soft photon mass (IR regulator), $\tilde{s}%
=s-m^{2}$,$~\tilde{u}=u-m^{2}$ and Li$(z)$ is the Spence function
(dilogarithm) defined by%
\begin{equation}
\text{Li}(z)=-\int_{0}^{z}dt\frac{\ln(1-t)}{t}.
\end{equation}
Using Eqs.(\ref{sigmaR}), (\ref{dltMT})  the $rhs$ in  Eq.(\ref{sgmRfin}) can be written as
\begin{equation}
\sigma_{R}^{1\gamma,\text{MT}}=G_{M}^{2}+\frac{\varepsilon}{\tau}G_{E}%
^{2}+2G_{M}\operatorname{Re}\left[  \delta\tilde{G}_{M}^{2\gamma}%
+\varepsilon\frac{\nu}{m^{2}}\tilde{F}_{3}-G_{M}\frac{1}{2}\delta_{2\gamma
}^{\text{MT}}\right]  . \label{sgmMT}%
\end{equation}

Let's look at the expression in square brackets in more detail. Substituting 
the explicit expressions for the amplitudes we obtain
\bea
\delta\tilde{G}_{M}^{2\gamma}+\varepsilon\frac{\nu}{m^{2}}\tilde{F}_{3}%
-G_{M}\frac{1}{2}\delta_{2\gamma}^{\text{MT}}&=&
-\frac{\alpha}{\pi}\left(\mathcal{F}_{1}(Q^2)\ln\frac{s}{\mu_{F}^{2}}\ln\left\vert \frac{s}{u}\right\vert
-G_{M}(Q^2)\ln\frac{\tilde{s}}{\mu_{F}^{2}}\ln\left\vert \frac{\tilde{s}}{\tilde
{u}}\right\vert \right)  \label{1st}%
\\
&+&\frac{\alpha}{\pi}\left[  \bar{C}_{M}(\varepsilon,Q^{2})+\varepsilon\frac
{\nu}{s}C_{3}(\varepsilon,Q^{2})\right] \mathcal{F}_{1}(Q^{2}) - G_{M}\frac{1}{2}\bar{\delta}_{2\gamma}^{\text{MT}}
 \label{2nd}
\\
&+&
\delta\tilde{G}_{M}^{(h)}(\varepsilon,Q^{2})+\varepsilon\frac{\nu}{m^{2}%
}\tilde{F}_{3}^{(h)}(\varepsilon,Q^{2}).  %
\label{3d}
\eea

From the first line (\ref{1st}) it follows that the  IR-regulator $\lambda^{2}$  cancels in the combination of the logarithms as it should be
\begin{equation}
g_{1}-G_{M}\frac{1}{2}\delta_{2\gamma}^{\text{MT}}=\frac{\alpha}{\pi}G_{M}%
\ln\frac{\tilde{s}}{\mu_{F}^{2}}\ln\left\vert \frac{\tilde{s}}{\tilde{u}%
}\right\vert -G_{M}\frac{1}{2}\bar{\delta}_{2\gamma}^{\text{MT}}.
\end{equation}
 This gives logarithm $\sim G_{M}\ln(\tilde{s}/\mu_{F}^{2}) \ln\vert \tilde{s}/\tilde{u}\vert $ with $s/\mu^2_F\gg 1$. This large logarithm
is proportional to magnetic FF $G_{M}$, which reflects that this contribution
arises from the hard-soft photon configuration. The scale $\mu_{F}\sim \Lambda$
plays the role of  UV cutoff for the soft photon virtuality. The
one more large logarithm also appears from the hard-hard photon contribution, see Eq.(\ref{GMF3s}). But  its coefficient  is proportional to WACS  FF $\mathcal{F}_{1}$.   Notice, that the coefficients of large logarithms in  Eq.(\ref{1st}) vanishes in the forward limit  (\ref{fwd}), which is provided by  the factor $\ln(-u/s)\simeq \ln(1+t/s)\sim t/s$ at large energy $s\to\infty$.

The hard-hard photon domain in the soft-spectator contribution also gives the term
$(\bar{C}_{M}+\varepsilon\frac{\nu }{s}C_{3})$  in Eq.(\ref{2nd}).
 This combination of the hard  functions also vanishes in the forward limit (\ref{fwd}).  
 Fulfilment of this condition provides an important check of the obtained analytical expressions. On the other hand this 
leads to a strong numerical cancellation between the  large numerical  terms in $\bar{C}_{M}$ and in 
$\varepsilon\frac{\nu}{s}C_{3}$.

The hard-spectator contributions $\delta\tilde{G}_{M}^{(h)}$ and $\frac{\nu
}{m^{2}}\tilde{F}_{3}^{(h)}$ in (\ref{3d})  independently vanish in the forward  limit
 which follows from  the structure of the perturbative kernels. Therefore the  boundary
condition (\ref{fwd}) is well satisfied. From this observation it follows that the simple 
extrapolation of Eq.(\ref{sgmMT} ) in the forward limit gives
\begin{equation}
\sigma_{R}^{1\gamma,\text{MT}}(\varepsilon=1,Q^{2})=G_{M}^{2}+\frac{1}{\tau
}~G_{E}^{2}= G_{M}^{2}\left(  1+\frac{\varepsilon}{\tau}R^{2}\right), \label{sgmfwd}
\end{equation}
where we introduced  the ratio $R=G_{E}/G_{M}$.
Let us  rewrite the Eq. (\ref{sgmMT})  as
\begin{equation}
\sigma_{R}^{1\gamma,\text{MT}}=G_{M}^{2}\left(  1+\frac{\varepsilon}{\tau}R^{2}\right)  +2G_{M}\operatorname{Re}\left[  \delta\tilde{G}_{M}^{2\gamma
}+\varepsilon\frac{\nu}{m^{2}}\tilde{F}_{3}-G_{M}\frac{1}{2}\delta_{2\gamma
}^{\text{MT}}\right]  , \label{sgmMT-R}%
\end{equation}
where we assume that  the ratio $R$  is fixed from the experimental  data  \cite{Jones00,Gayou02,Punjabi:2005wq,Puckett:2010ac}. 
Then the discrepancy between polarised and  unpolarised data  can  be  explained by the  TPE contribution in Eq.(\ref{sgmMT-R}).  
Remind,  that obtained expressions for the TPE amplitudes are only valid in the region where\emph{ all}
Mandelstam variables are sufficiently large (\ref{def:hard}).
 This means that we can not  describe the backward region where variable  $u$ is
small. The minimum value of $u_{\min}$ corresponding to the boundary $u>u_{\min}$ is equivalent to the condition $\varepsilon>\varepsilon_{\min}$ and will be considered further.

Many existing  data allows one to conclude that  behavior of the  reduced cross section $\sigma_{R}^{1\gamma,\text{MT}}(\varepsilon,Q^{2})$ at relatively low values of $Q^{2}$ can be sufficiently well described by the  linear empirical fit
\begin{equation}
\sigma_{R}^{1\gamma,\text{MT}}(\varepsilon,Q^{2})=a+\varepsilon b.
\label{sgmlin}%
\end{equation}
For higher values of $Q^2$,
 it is assumed that this behaviour also holds for the entire interval $0<\varepsilon<1$ at fixed $Q^{2}$ that suggests that  the nonlinear effects in TPE are very
small. If this is so, then the empirical constants $a$ and $b$ can be obtained from the data corresponding to the given cross section at fixed $Q^{2}$.  Combining (\ref{sgmfwd})
and (\ref{sgmlin}) one finds \cite{Guttmann:2010au}
\begin{equation}
G_{M}^{2}=\frac{a+b}{1+R^{2}/\tau}.
\end{equation}
The value of $G_{M}^{2}$ thus obtained, used in equation (\ref{sgmMT-R}), ensures that the linear behaviour (\ref{sgmlin}) does not contradict to the boundary condition (\ref{sgmfwd}).

\section{Estimates of TPE effect  in  elastic electron-proton scattering}
\label{proton}

In order to  perform numerical estimates let us  fix the non-perturbative
input in the TPE amplitudes. For the FF  $\mathcal{F}_{1}$ we use the
simple empirical parametrisation which was obtained in Ref.\cite{Kivel:2015vwa}
\begin{equation}
\mathcal{F}_{1}(Q^{2})=\left(  \frac{\Lambda^{2}}{Q^{2}}\right)  ^{\alpha},
\label{Fwacs}%
\end{equation}
where  $\Lambda=1.174$GeV$^{2}$ and the exponent $\alpha
=2.091$. This FF provides a good description of the measured WACS cross section
for different energies $s$  in the interval $2.5\,$GeV$^{2}<Q^{2}<6.5\,$GeV$^{2}$  and $-u\geq
2.5\,$GeV$^{2}$. This restriction can be easily converted to the
boundary for the photon polarisation  $\varepsilon_{\min}<\varepsilon < 1$ for fixed value of $Q^{2}$.
 For the value of the factorisation scale in Eq.(\ref{1st}) we take $\mu_{F}=200\,$MeV.

In order to describe the hard spectator terms we  take the model for the nucleon distribution amplitude (DA)  from Ref.\cite{Anikin:2013aka}
\bea
\varphi_{3}(x_{1},x_{2},x_{3})&=&120\, f_N\, x_{1}x_{2}x_{3}\, \bar \varphi_{3}(x_i, \mu), \label{phi3}\\
\bar \varphi_{3}(x_i)&=&  1+\sum
_{j=0}^{1}\phi_{1j}(\mu)P_{1j}(x_{1},x_{2},x_{3})+\sum_{j=0}^{2}\phi_{2j}%
(\mu)P_{2j}(x_{1},x_{2},x_{3}) ,
\eea
where $P_{ij}(x_i)$ are homogenous orthogonal polynomials
\bea
P_{10}(x_i)&=&21(x_1-x_3),\ \ P_{11}(x_i)=7(x_1-2x_2+x_3),\\
  P_{20}(x_i)&=&\frac{63}{10}[ 3(x_1-x_3)^2-3x_2(x_1+x_3)+2x_2^2 ],  \\ 
  P_{21}(x_i)&=&\frac{63}{2} (x_1 -3 x_2+x_3)(x_1-x_3),\\
  P_{22}(x_i)&=&\frac{9}{5}[ x_1^2+9x_2(x_1+x_3)-12x_1x_3-6x_2^2+x_3^2].
\eea
 The moments $\phi_{ij}(\mu)$ are multiplicatively renormalisable,  see more  details in Refs. \cite{Anikin:2013aka, Braun:2008ia}.
Their values are estimated with the help of light-cone sum rules for the electromagnetic FF's.  For our analysis we will use the ABO1 model \cite{Anikin:2013aka} and the COZ  model from Ref.\cite{Chernyak:1987nv}.  The numerical values for the coefficients $\phi_{ij}$ are given in Table~\ref{LCDAmodels}. 
 For the  normalisation constant $f_{N}$ in (\ref{phi3})  we use 
\begin{equation}
f_{N}=\left(  5.0\pm0.5\right)  \times10^{-3}\text{GeV}^{2},
\end{equation}
obtained from the QCD sum rules \cite{Chernyak:1987nv}.  

\begin{table}[th]
\centering
\begin{tabular}{|c|c|c|c|c|c| } 
\hline
model &  $\varphi _{10}$ & $\varphi _{11}$ & $\varphi _{20}$ & $\varphi _{21}$ & $\varphi _{22}$ 
 \\ \hline
ABO  & $0.050$ & $0.050$ & $0.075$ & $-0.027$ & $0.170$ 
 \\ \hline
COZ  & $0.154$ & $0.182$ & $0.380$ & $0.054$ & $-0.146$
 \\ \hline
\end{tabular}
\caption{ The numerical values of DA parameters $\phi_{ij}$ at  $\mu^{2}=4\text{~GeV}^{2}$ .}  
\label{LCDAmodels}
\end{table}

The scale of the running coupling $\alpha_{s}(\mu_{R}^{2})$ is set to be
$\mu_{R}^{2}\simeq0.6~Q^{2}$ and we use the two-loop running coupling with
$\alpha_{s}(1.5)=0.360$.

In our numerical calculations we also use the ratio $R=G_{E}/G_{M}$ which is
fixed from the experimental data  \cite{Jones00,Punjabi:2005wq, Gayou02,Puckett:2010ac}. For
simplicity,  we perform  the following empirical fit of the data for  the
interval $0.5-8.5$~GeV$^{2}$ in $Q^{2}$:
\begin{equation}
R(Q^{2})=\mu_{p}^{-1}(1-a\ln^{2}[Q^{2}/\Lambda
_{R}^{2}]), \label{def:R}%
\end{equation}
with $a=0.070\pm0.007$, $\Lambda_{R}=0.54\pm0.03$~GeV. The other parameters
read  $m=938\text{ MeV},~\mu_{p}=2.7928$.

\begin{figure}[ptb]
\centering
\includegraphics[height=1.8713in,width=2.983in]
{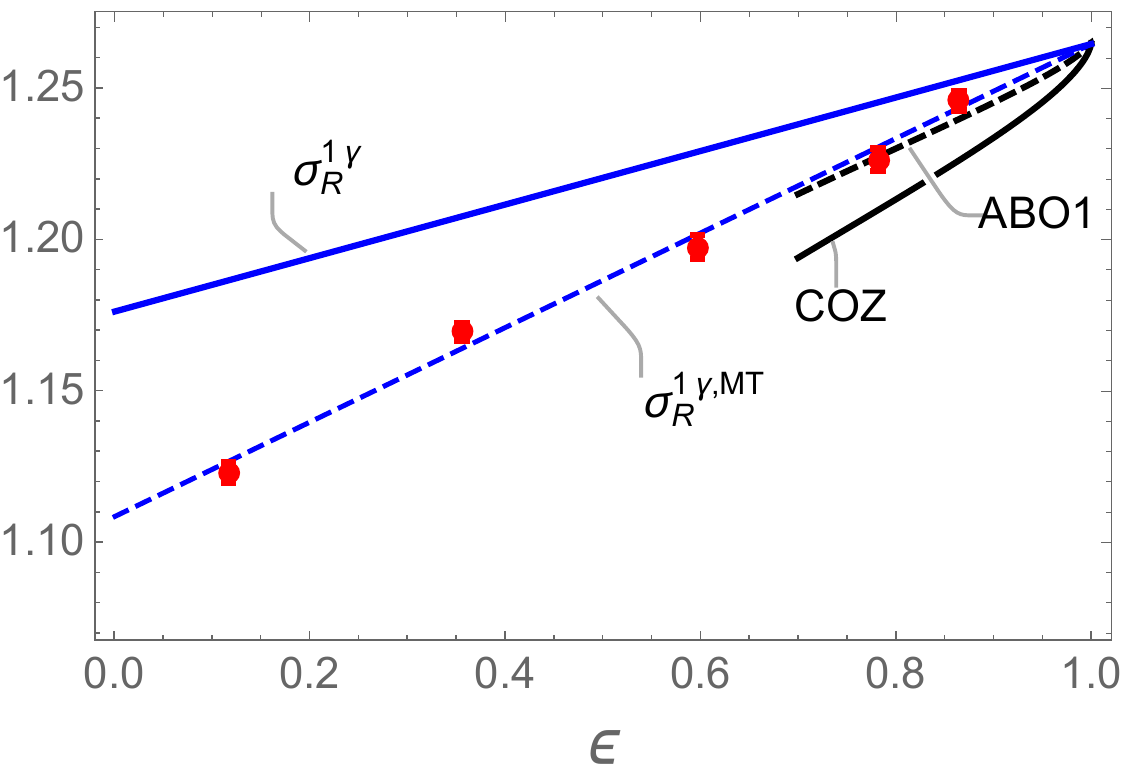}
\includegraphics[height=1.8713in,width=2.983in]
{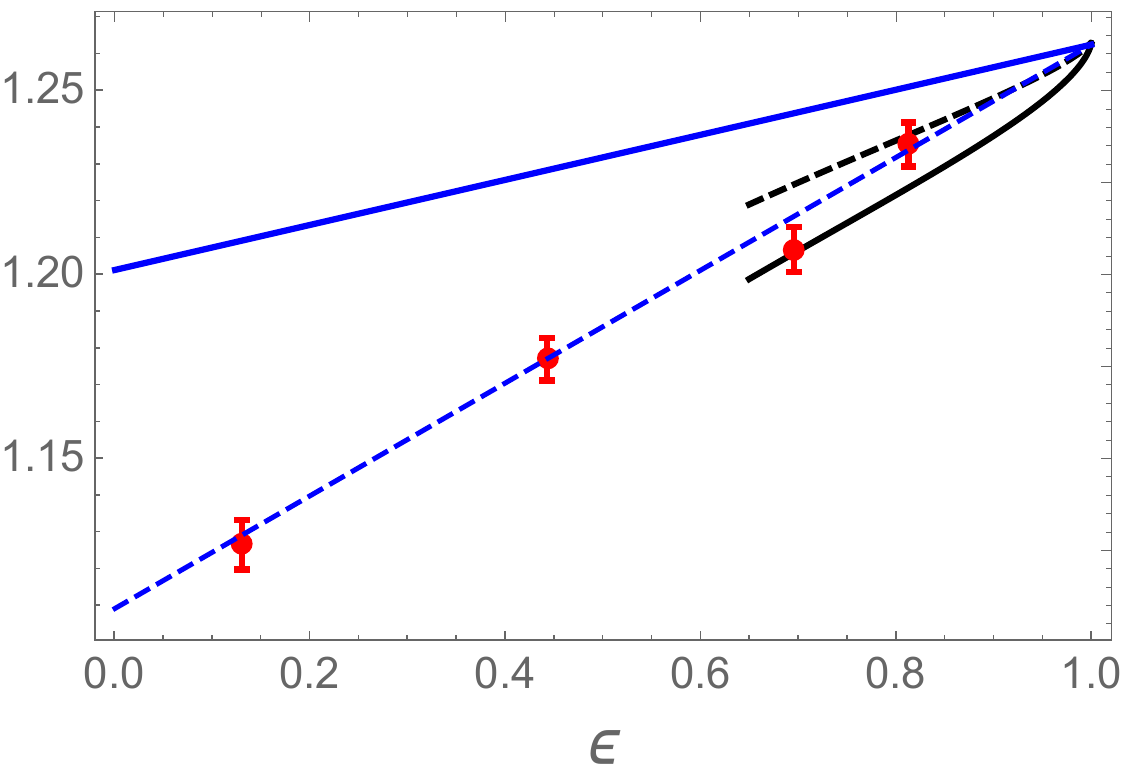}
\caption{The reduced cross section as a function of $\varepsilon$ at fixed
values of $Q^{2}=2.64,3.20\ $GeV$^{2}$, the data are from  Ref. \cite{Qattan:2005zd}.}
\label{tpe264}
\end{figure}

\begin{figure}[ptb]
\centering
\includegraphics[height=1.8713in,width=2.983in]
{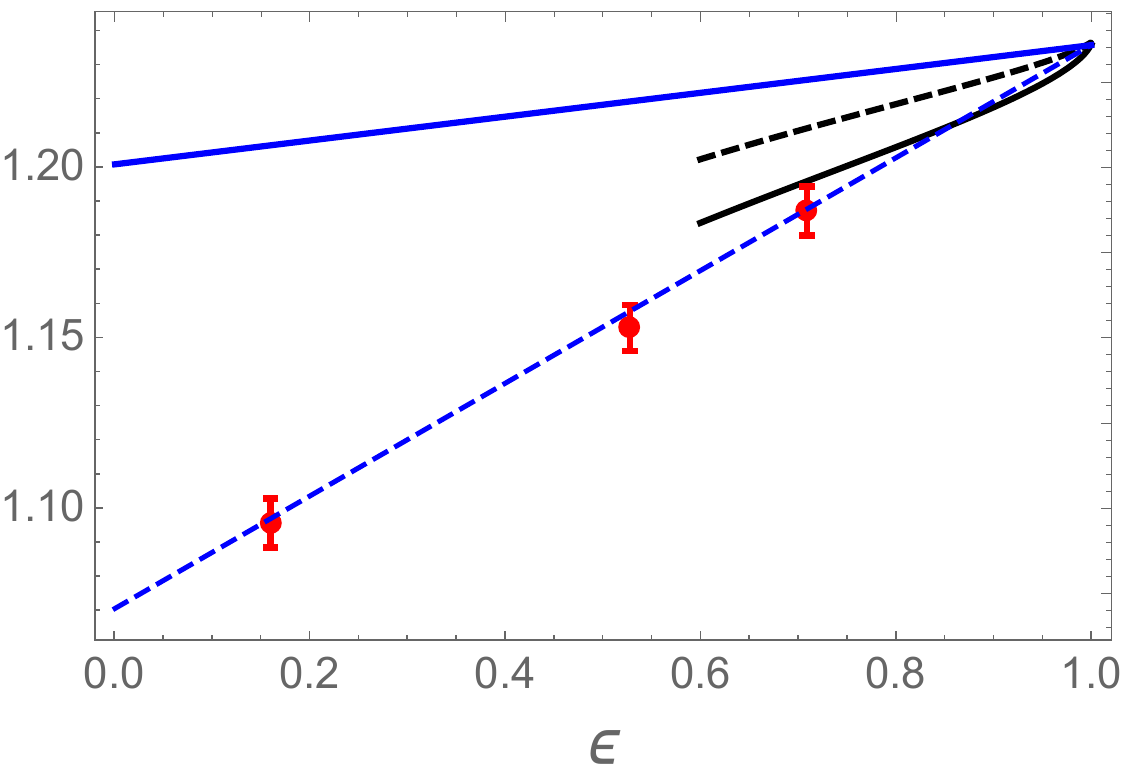}
\includegraphics[height=1.8713in,width=2.983in]
{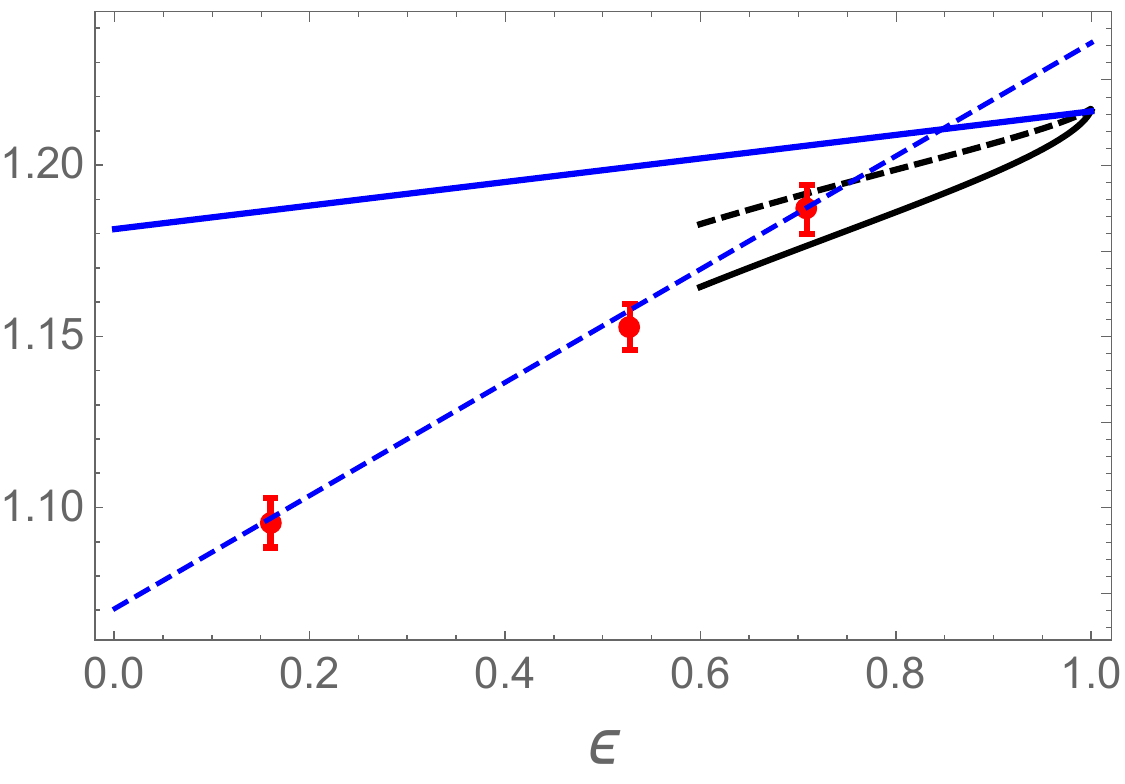}
\caption{Left chart: the reduced cross section as a function of $\varepsilon$ at fixed
values of $Q^{2}=4.10\ $GeV$^{2}$, the data are from  Ref. \cite{Qattan:2005zd, Qattan:2024pco}. 
Right chart: the same but assuming nonlinear behaviour and smaller FF $G_M$, see more details in the text. }
\label{tpe410}
\end{figure}

Combining  this input with the linear fit (\ref{sgmlin}), using $G_{M}$ as in Eq.(\ref{sgmMT-R})  the analysis of the  JLab data from  Ref.\cite{Qattan:2005zd}
gives  results, which are shown in Fig.$\,$\ref{tpe264} and  Fig.$\,$\ref{tpe410}.

The blue solid line in these figures shows the reduced cross section with the slope obtained
from the polarisation transfer experiments.  For a convenience the cross section is normalised to the square of the dipole FF $G_{dip}(Q^2)=\mu_p/(1+Q^2/\Lambda^2)^2$, with $\Lambda=0.71\,$GeV$^2$.  The blue dashed line corresponds to the linear fit  of the unpolarised data for $\sigma_{R}^{1\gamma,\text{MT}}$  (red points).  

These lines are extrapolated to the point $\varepsilon=0$ where the TPE effect must vanish. The different slopes  of these lines  must be explained.   

  The  black solid and dashed lines represent the reduced cross section with the TPE
contribution for COZ and ABO1 model, respectively. These lines are only
shown in the region $\varepsilon>\varepsilon_{\min}$ where our approximation is valid.

The numerical value of the TPE effect is sensitive to the choice of the model for nucleon DA.   
The obtained plots show that TPE with ABO1 model of DA works reasonably well for
$Q^{2}=2.6 \text{ and } 3.2\,$GeV$^{2}$ but  for $Q^{2}=4.10\,$GeV$^{2}$  the corresponding numerical
effect is already somewhat small.  For  COZ model gives large numerical effect: 
large effect at $Q^{2}=2.6\,$GeV$^{2}$ and quite reliable
description for  $Q^{2}=3.2\text{ and }4.10\,$GeV$^{2}$.  Such a result  illustrates that the obtained
description is sensitive to the hard-spectator contribution, which
turns out to be  quite large comparing to the soft-spectator one.

\begin{figure}[ptb]
\centering
\includegraphics[height=1.8713in,width=2.9913in]{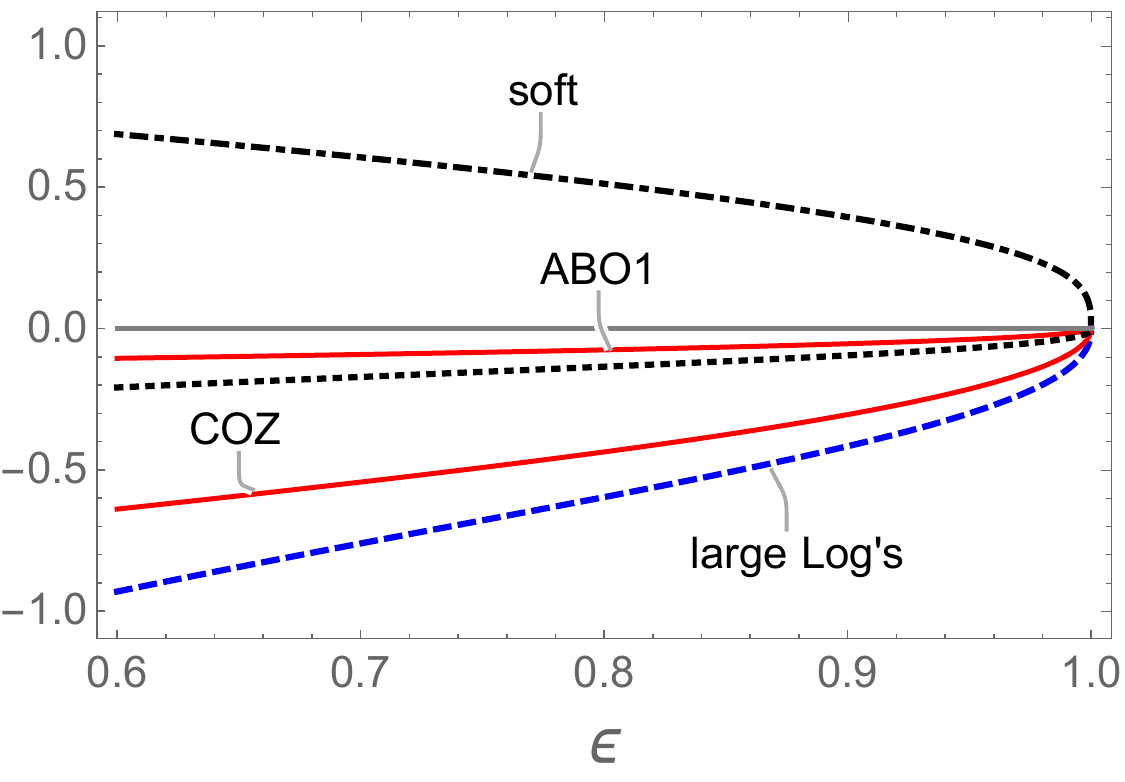}
\caption{The various  TPE contributions as shown in Eqs.(\ref{1st})-(\ref{3d}) at fixed $Q^{2}=4.10\,$GeV$^{2}$. See further explanation in the text.}
\label{individual}
\end{figure}

 In order to better understand this point,  we plot in Fig.$\,$\ref{individual}
 the different contributions for the TPE correction in Eq.(\ref{sgmMT-R})
 for $Q^{2}=4.10\,$GeV$^2$. The TPE contributions are represented as a sum of
 four terms according to Eqs.(\ref{1st})-(\ref{3d}).
  For a convenience
we normalise  the TPE corrections to the value $R^{2}/\tau$. 
The  largest contributions are generated by the logarithms in
Eq.(\ref{1st}) (blue dashed line with the label ``large Log's'') and by  the soft-spectator contribution $\left(  \bar{C}%
_{M}+\varepsilon\frac{\nu}{s}C_{3}\right) \mathcal{F}_{1}$ (black,
dot-dashed line with the label ``soft"), which have opposite sign.  The hard-spectator term $ \delta\tilde{G}_{M}
^{(h)}+\varepsilon\frac{\nu}{m^{2}}\tilde{F}_{3}^{(h)}$ for the different models  are shown by solid (red) curves with the labels ``COZ"  and  ``ABO1", respectively.  
The MT subtraction term $G_{M}\bar{\delta}_{2\gamma}^{\text{MT}}/2$ is shown by black dotted curve.  

The soft-spectator contribution is positive while the combination with large logarithms gives the negative contribution. Together these terms give about $-30\%$ effect  (relative to $R^{2}/\tau$) at $\varepsilon_{\min}=0.6$. The MT subtruction term  gives additional $-10\%$. However,  the resulting correction is still  small and  in
case of ABO1 model such a description gives  only a half of required effect. The
hard-spectator contribution with the COZ model gives a much larger numerical effect, 
which is about factor six larger comparing to ABO1 model and gives about
$-60\%$.  This allows one to describe the gap between the  polarised and
 unpolarised linear curves in Fig.$\,$\ref{tpe410}.  

The strong cancellation between the soft-spectator  (\ref{2nd}) and large logarithm (\ref{1st}) contributions, 
which makes the overall result to be  more sensitive to the hard spectator
term (\ref{3d}), is not accidental.  The large positive contribution occurs
in the combination of the hard functions $\bar{C}_{M}+\varepsilon\frac{\nu}
{s}C_{3}$ and is dominated by the simple numerical constant
\begin{equation}
\bar{C}_{M}+\varepsilon\frac{\nu}{s}C_{3}=(1-\varepsilon)\frac{\pi}%
{2}+\mathcal{O}(s^{-1})~.
\end{equation}
This  behaviour  ensures the fulfilment of the boundary
condition (\ref{fwd}).  This is important property of the box diagram in the hard region.

\begin{figure}[ptb]%
\centering
\includegraphics[height=4.11cm,width=6.5519cm]
{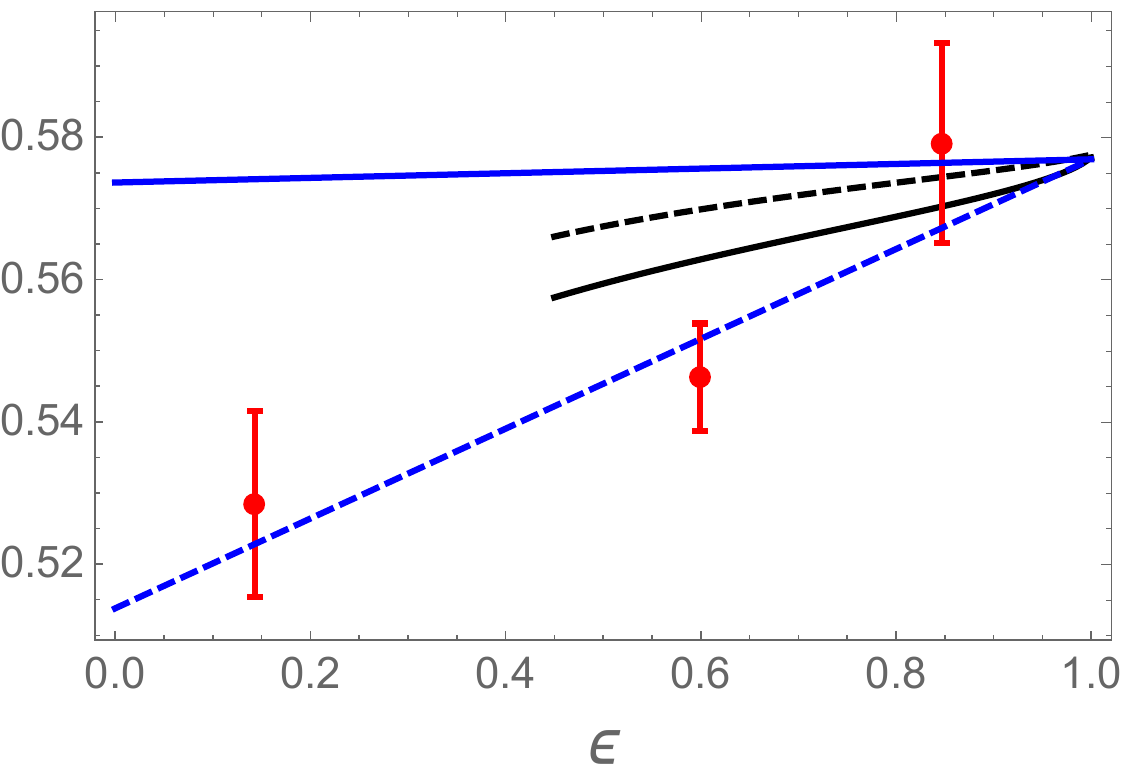}
\includegraphics[height=4.11cm,width=6.5519cm]{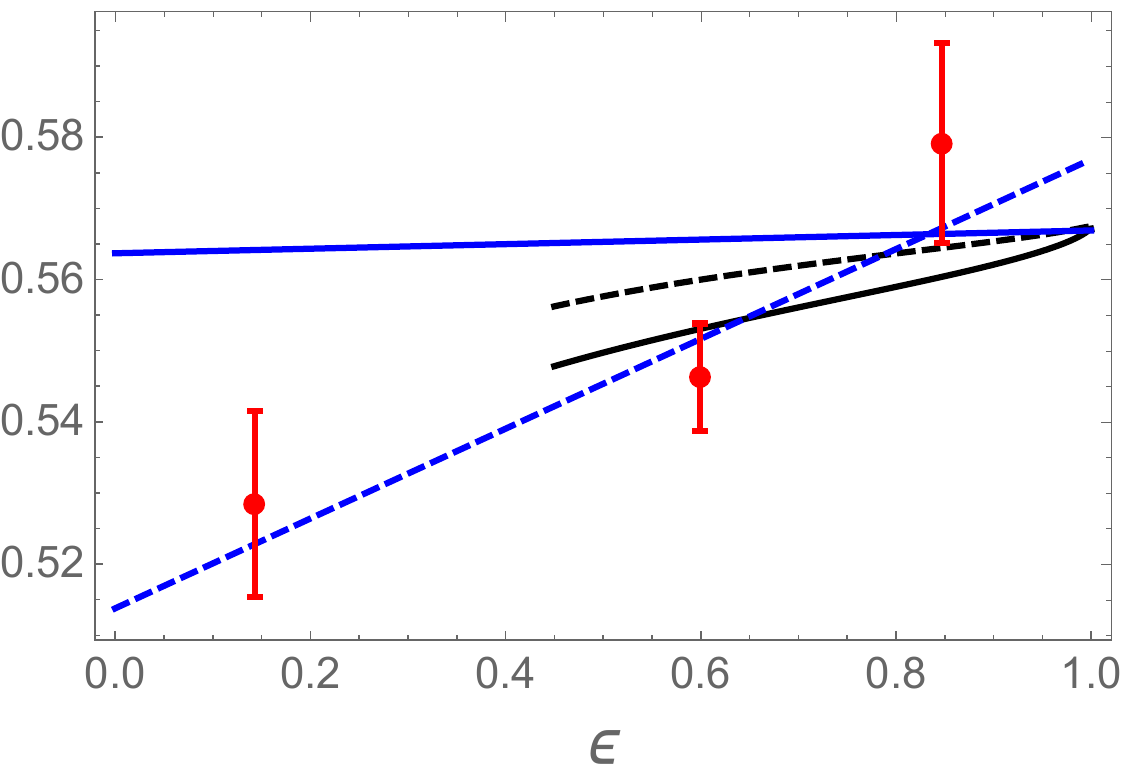}
\label{tpe700mod}
\caption{Left chart: the same as in Fig.\ref{tpe410},  $Q^{2}=7\ $GeV$^{2}$, the data are from
Ref.\cite{Christy:2021snt, Andivahis:1994rq} . Right chart: the same but assuming nonlinear behaviour and smaller FF $G_M$. }
\label{tpe700}
\end{figure}

Our numerical estimates show that obtained TPE effect decreases with increasing $Q^{2}$.
  However,  the  existing data \cite{Christy:2021snt,Andivahis:1994rq,Sill:1992qw} 
indicate that TPE effect remains quite large  at large values of $Q^2$. In order to illustrate this  we show  in Fig.$\,$\ref{tpe700}  the same plots as in Fig.$\,$\ref{tpe410}   
but for $Q^{2}=7\,$GeV$^2$.
Notice, that in this case we need to extrapolate the empirical formula
(\ref{Fwacs}) for the FF $\mathcal{F}_{1}$, which also provides additional ambiguity.  
Nevertheless, from Fig.$\,$\ref{tpe700} it is clearly seen that even with the COZ model  it is not possible to describe the difference between the different slopes. 

This behaviour may indicate that there is an effect beyond the description developed that needs to be taken into account.  This effect is especially important for sufficiently large values of  $Q^{2}>4$GeV$^{2}$. 

One possible solution is  that assumption about the linear  $\varepsilon$-behaviour of the
reduced cross section on the whole interval must be  revised.There are no strict theoretical arguments in favor of the fact that the TPE amplitudes should have a linear behavior in $\varepsilon$ over the entire interval.
For example, the slope in the region $\varepsilon>\varepsilon_{min}$ for large $Q^{2}$ may differ from the slope in the region $\varepsilon<\varepsilon_{min}$. The reason for this behaviour may be that the TPE amplitudes depend on completely different underlying scattering mechanisms for small and large values of   $\varepsilon$ (backward
and forward domains, respectively). Then such an effect can play an important role at large values of $Q^{2}$.
The absolute value of such an effect may be relatively small and therefore may not have a strong impact on the value of $G_{M}$, but may play an important role in understanding the FF ratio discrepancy.

In such case  the value of the FF $G_M$ can not be obtained by the simple   
extrapolation of the linear behaviour  to the point $\varepsilon=0$ or $\varepsilon=1$. 
%because of TPE effects
%\bea
%\sigma_{R}^{1\gamma,\text{MT}}(\varepsilon=0, Q^2)=G_{M}^{2}(Q^2)  +2G_{M}(Q^2)\operatorname{Re}
%\left[  
%\delta\tilde{G}_{M}^{2\gamma}(\varepsilon=0, Q^2)-G_{M}(Q^2)\frac{1}{2}\delta_{2\gamma}^{\text{MT}}(\varepsilon=0, Q^2)
%\right].
% \label{sgmMT-R0}%
%\eea
  In  order to illustrate corresponding effect let us consider a small modification of the
calculations in Fig.$\,$\ref{tpe410}  and  Fig.$\,$\ref{tpe700} and assume that the slope  in the region $\varepsilon>\varepsilon_{\min}$ is approximately agree with our calculations but the value of $G_M$ is smaller than one obtained from the linear fit  in the total region $0<\varepsilon<1$.  The results are
shown in the right plots of the same figures.  As a result the value of the FF $G_{M}$ was reduced by  $1.7\%$ and $1.3\%$ for $Q^{2}=4.10$ and $7.0$ GeV$^{2}$, respectively.  
It is evident that such behaviour allows us to qualitatively understand the discrepancy in the region $\varepsilon>\varepsilon_{\min}$.  This assumption can be verified experimentally by precise measurements of the reduced cross section in the region $\varepsilon>\varepsilon_{\min}$.

The TPE calculation under discussion involves various uncertainties. These uncertainties can be divided on theoretical and experimental. The latter  are associated with the data which are used  to estimate  non-perturbative  quantities, like nucleon DA, $\mathcal{F}_{1}$.  Theoretical ambiguities are related to the various QCD next-to-leading and power corrections, which can provide some effect at moderate values of $Q^2$. 

  Thus, one can see that  the TPE effect  is quite sensitive to the model of the proton DA.  The COZ model  gives much  larger numerical effect because this distribution is  sensitive to the configuration where the spectator quark momentum fraction is small. It is possible that this indicates a large contribution from soft overlap associated with the so-called "cat's ear" configuration, where photons interact with different quarks. Such a contribution has not yet been studied in the literature, since it is suppressed by the inverse power of $1/Q^2$.
  
The existing lattice and non-perturbative  results 
indicate that the value of the normalisation constant $f_{N}$  can be  smaller then
one used in this work, see {\it e.g.} Ref.\cite{Bali:2019ecy, Kim:2021zbz}.  Using such a small value of $f_{N}$ will further suppress the contribution of the hard observer (by about a factor of two), reducing the effect of TPE.  At the same time, smaller value of $f_{N}$ also reduces the pQCD (hard spectator contribution) predictions for nucleon FF and WACS, indicating a larger effect of the soft-spectator mechanism in these hard exclusive reactions.

\section{TPE correction in  elastic electron-neutron scattering}
\label{neutron}

A discussion of the TPE for the neutron target can also be important for a
Rosenbluth analysis of  data for reduced cross section at higher values of
$Q^{2}$. The main difference with the proton is that the TPE amplitude is
free from the QED IR-divergency. On the other hand the contribution with the
hard photons remains significant.

The theoretical description of the TPE amplitudes can be carried out in the same way
but now we expect that the amplitudes associated with the
soft photon are very small 
\begin{equation}
\delta_{2\gamma}^{\text{MT}}=0,~g_{1n}(\varepsilon,Q^{2})=0.
\end{equation}

From existing data \cite{Riordan:2010id}  one can conclude that at least in the
region $Q^{2}<4\,$GeV$^{2}$ the ratio $R_{n}=G_{En}/G_{Mn}$ is sufficiently
small and  the estimate in Eq.(\ref{GEsmall})  also works 
in this case.  Therefore Eq.(\ref{sgmMT}) can be used for the
phenomenological estimates.

The required amplitudes, which enter in the expression for  the reduced
cross section in (\ref{sgmMT}) now reads
\bea
\delta\tilde{G}_{Mn}^{2\gamma}+\varepsilon\frac{\nu}{m^{2}}\tilde{F}%
_{3n}&=&-\frac{\alpha}{\pi}\mathcal{F}_{1n}(Q^{2})\ln\frac{s}{\mu_{F}^{2}}%
\ln\left\vert \frac{s}{u}\right\vert 
\nonumber\\
&+&\frac{\alpha}{\pi}\left[  \bar{C}_{M}(\varepsilon,Q^{2})+\varepsilon\frac
{\nu}{s}C_{3}(\varepsilon,Q^{2})\right]  \mathcal{F}_{1n}(Q^{2})+\delta
\tilde{G}_{Mn}^{(h)}+\varepsilon\frac{\nu}{m^{2}}\tilde{F}_{3n}^{(h)},
\label{TPEn}%
\eea
where subscript $n$ denotes a neutron amplitudes and FF's. The first two terms in the {\it rhs} of Eq.(\ref{TPEn}) correspond to the
soft-spectator contribution. The large logarithm arises from the $\delta
\tilde{G}_{Mn}^{(s)}$. The hard coefficient functions $\bar{C}_{M}$ and
$C_{3}$ are the same as for the proton, see  Eqs.(\ref{CM}) and
(\ref{C3}).

The hard-spectator contributions can be described very similar to a proton
case, with the difference  due to the interchange of the quarks charges
$e_{u}\leftrightarrow e_{d}$ in the hard amplitudes. The nucleon DA is the same.

The soft-spectator contribution depends on the FF $\mathcal{F}_{1n}$ which is
different from the proton FF $\mathcal{F}_{1}$ due to the different quark
electromagnetic charges.  This FF can be obtained  from the WACS on the neutron like $\mathcal{F}_{1}$, but such measurements have not yet been
done. We  get an estimate of this quantity using the isotopic symmetry,  nucleon electromagneic FFs and WACS FF  $\mathcal{F}_{1}$ obtained for the proton.  
The current data allows one to get the values $\mathcal{F}_{1n}$ for $Q^{2}=2.5$ and
$3.4\,$GeV$^{2}$. The details are provided in Appendix$\,$\ref{appC}. These  calculations give
\begin{equation}
\mathcal{F}_{1n}(2.5)\simeq0.25~,~\ \mathcal{F}_{1n}(3.4)\simeq0.12.
\label{F1n-num}%
\end{equation}

The reduced cross section for a neutron target reads
\begin{equation}
\sigma_{Rn}=G_{Mn}^{2}\left(  1+\frac{\varepsilon}{\tau}R_{n}%
^{2}\right)  +2G_{Mn}\operatorname{Re}\left[  \delta\tilde{G}_{Mn}^{2\gamma
}+\varepsilon\frac{\nu}{m^{2}}\tilde{F}_{3n}\right] ,
\end{equation}

The value of the ratio $R_{n}$  we take from the data in Ref.\cite{Riordan:2010id}
\begin{equation}
\mu_{n}R_{n}(2.5)=0.397~,~\ \ ~\ \mu_{n}R_{n}(3.4)=0.481,
\label{ref:Rn}
\end{equation}
where the neutron  magnetic moment $\mu_{n}=-1.913$.  The values of the  FF
$G_{Mn}$ we take from Ref.\cite{Lachniet:2008qf}
\begin{equation}
G_{Mn}(2.5)=0.0506~\mu_{n},~\ ~G_{Mn}(3.4)=0.0296~\mu_{n}.
\end{equation}

\begin{figure}[ptb]
\centering
\includegraphics[width=5.0in]
{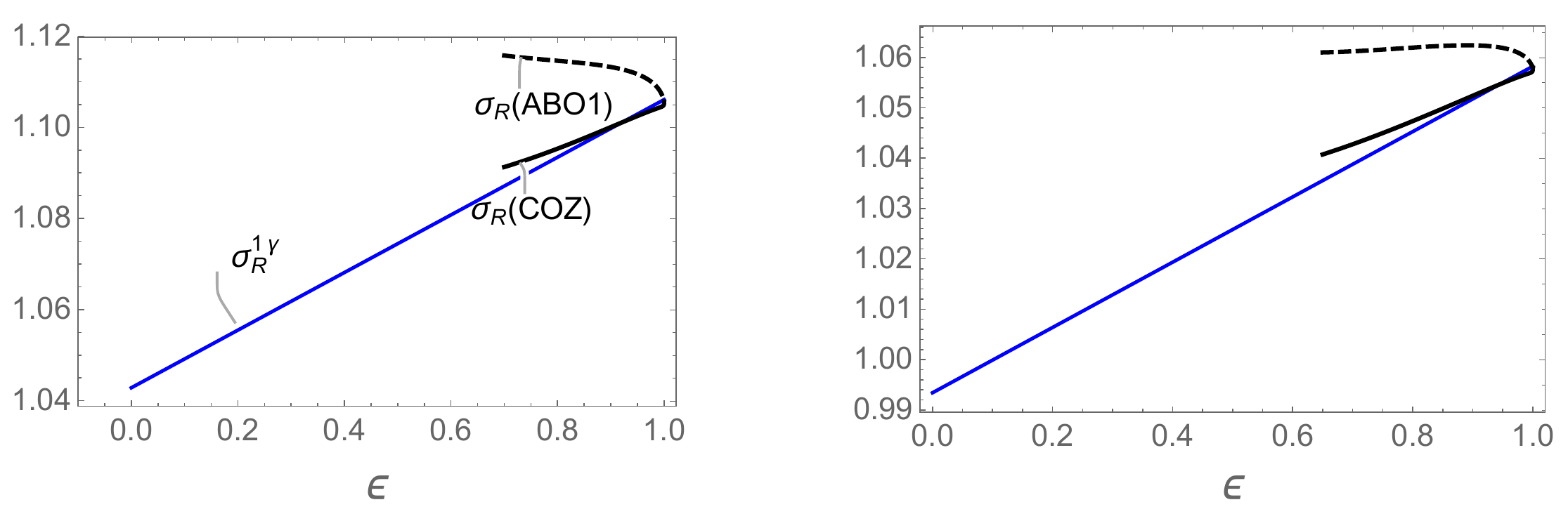}
\caption{The normalised reduced cross section $\sigma_{R}/(\mu_{n}G_{dip})^{2}$ as a
function of $\varepsilon$ at fixed values of $Q^{2}=2.5$ (left chart) and $3.4\ $GeV$^{2}$ (right chart).
The solid blue curve corresponds to the one-$\gamma$ exchange with the ratios $R_n$ from Eq.(\ref{ref:Rn}).  
The black dashed and solid line describe the cross section with the different TPE  contributions. }
\label{tpe-n}
\end{figure}

The obtained TPE effect is shown in Fig.$\ $\ref{tpe-n}. We  show the results 
in the region $\varepsilon>\varepsilon_{\min}$ only.
The solid blue line gives the reduced cross section without the TPE
contribution. The black dashed and solid lines describe the reduced cross
section with the TPE effect for the ABO1- and COZ-models of the nucleon DA, respectively.  The
big difference for the different models is explained by the cancellation
between the soft-spectator (which is positive) and the hard-spectator
contribution. Taking into account  the experience with the proton cross section we expect that
ABO1-model provides a more realistic description. In this case the
hard-spectator contribution is relatively small and  obtained numerical effect is about $2.2-2.5\%$ relative to
the Born cross section. Such a correction is comparable with the one in the
proton case but  it has the opposite sign. This reduces the slope,
 providing  the smaller ratio $R_{n}$. The obtained estimates are applicable for a small region $\varepsilon>\varepsilon_{min}$, however, it can be assumed that the TPE effect will be even greater in the region of large $\varepsilon$, as follows from the estimates of the hadron model \cite{Blunden:2005ew}. Therefore, the TPE correction for the neutron cross section is also important and should be taken into account when analysing the experimental data.

\section{Discussion}
\lab{disc}

Motivated by measurements at large $Q^2$, we re-analyse the TPE effect in elastic electron-proton scattering.  We use the effective field theory formalism  developed in Ref.\cite{Kivel:2012vs}.    This description is different from the existing hadronic and dispersive approaches because it  systematically separates and includes the short and long distance effects. The short distance amplitudes are computed in pQCD,  the long distance quantities are restricted using different experimental data or a low  energy EFT.  The resulting description works only in the region where all kinematic variables are large (\ref{def:hard}). In other words, for a fixed $Q^2$ the formalism is applicable only in a certain region $\varepsilon>\varepsilon_{min}$. The value of $\varepsilon_{min}$ depends on the energy and momentum transfer and  $\varepsilon_{min}\to 0$ as $Q^2\to \infty$. 

The developed formalism takes into account the soft- and hard-spectator contributions, which are associated with the different underlying hard scattering of the quarks.  It is shown that there are strong cancellations between different large terms in the TPE amplitudes.  It is also found that at large $Q^2$ the contribution of excited hadronic states is strongly suppressed in the region where one photon is soft.  As a result   the value of the TPE effect is quite sensitive to the so-called  "cat ear" diagrams, which have been computed  in the hard spectator approximation.   We perform a numerical analysis using two different models for the nucleon distribution amplitude.    
  
  The developed description allows us to explain the discrepancy between the FF data using TPE for $Q^2=2.5-3.5\,$GeV$^2$, but for larger $Q^2$ the obtained TPE correction is no longer large enough, provided that the cross section depends linearly on the photon polarisation. Probably, the basic assumption about the linear behaviour of the reduced cross section with respect to $\varepsilon$ over the entire interval for fixed $Q^2$ should be critically revised for large values of $Q^2$. The different behavior can be naturally explained by the different scattering dynamics in the region of small $\varepsilon<\varepsilon_{min}$ and large $\varepsilon>\varepsilon_{min}$ photon polarisation. 
  
  We show that using a different slope in the region $\varepsilon>\varepsilon_{min}$ changes the extracted $G_M$ value by $1-2\%$ for $Q^2\leq 7\,$GeV$^2$, but they also have a strong effect on the extracted small $G_E/G_M$ ratio. Note also that some nonlinearities were also observed in the dispersion calculations of $Q^2<5\,$GeV$^2$ \cite{Ahmed:2020uso}. 
  
  The developed formalism have been used to estimate the TPE effect in the electron-neutron scattering. In this case the long distance effects associated with the soft photon interaction absent  and the dominant  contribution is only associated with the hard photons. However,  the  specific hard-collinear  FF $\mathcal{F}_{1n}$, which is defined within the SCET framework and describes the soft-spectator scattering,  can not  be  restricted from the data as in case of  proton target. Therefore we use isospin symmetry and existing data for the proton and neutron electromagnetic FFs in order to estimate this quantity.   Our estimates for $Q^2=2.5 - 3.5\,$GeV$^2$ allow us to conclude that the obtained TPE effect is quite large and comparable to that in the proton cross section.

\section*{\Huge Appendix}
\appendix
\numberwithin{equation}{section}
\setcounter{equation}{0}
\section{Analytical expressions for the  soft- and hard-spectator  TPE amplitudes}
\label{appA}
For a convenience, we provide the analytical expressions for the amplitudes
used in the paper.

\bea
\delta\tilde{G}_{M}^{(s)}(\varepsilon,Q^{2})+\varepsilon\frac{\nu}{m^{2}%
}\tilde{F}_{3}^{(s)}&=&\frac{\alpha}{\pi}~g_{1}(\varepsilon,Q,\mu_{F})+\frac{\alpha}{\pi}\ln\left\vert \frac{u}{s}\right\vert \mathcal{F}_{1}(Q^{2})
\ln\frac{s}{\mu_{F}^{2}} \nonumber \\
& +&\frac{\alpha}{\pi}\left\{  \bar C_{M}%
(\varepsilon,Q^{2})+\frac{\nu}{s}C_{3}(\varepsilon,Q^{2})\right\}
\mathcal{F}_{1}(Q^{2}),
\label{GMF3s}
\eea
The hard coefficient functions read
\begin{align}
\bar C_{M}(\varepsilon,Q^{2})   =
\frac{1}{2}\ln^{2}\bar{z}-\frac{1}{4}\frac{z\ln^{2}%
z}{1-z}-\frac{z\ln^{2}\bar{z}/z}{4}-\frac{1}{2}\ln\bar{z}+\frac{2-z}
{4}\pi^{2}.
\label{CM}
\end{align}
\begin{equation}
C_{3}(z,Q^{2})=\frac{2-z}{2\bar{z}^{2}}\ln^{2}z-\frac{2-z}{2}\ln^{2}\frac
{\bar{z}}{z}+\frac{1}{\bar{z}}\ln z+\ln\frac{\bar{z}}{z}-\frac{2-z}{2}\pi
^{2}.\label{C3}
\end{equation}
where $z=-t/s>0$.
These expressions are in agreement with the analogous calculations in Ref.\cite{Afanasev:2005mp}.

The hard-spectator contribution for the proton case reads\footnote{These expressions have factor additional $1/2$ comparing to ones in Ref.\cite{Kivel:2009eg}, where this factor was overlooked. }
\bea
\delta\tilde{G}^{(h)}_{M}=-\frac{~~\alpha\,\alpha_{s}(\mu^{2})}{Q^{4}}\left(
\frac{4\pi}{3!}\right)  ^{2}\,\frac{1}{2}(2\zeta-1)~J_{M}(\zeta),
\\
\frac{\nu}{M^{2}}\tilde{F}^{(h)}_{3}=-\frac{~~\alpha\, \alpha_{s}(\mu^{2})}{Q^{4}%
}\left(  \frac{4\pi}{3!}\right)  ^{2}\,\frac{1}{2}(2\zeta-1)~J_{3}(\zeta),
\eea
where $\zeta=1/z=s/(-t)>1$.  The convolution integrals read
\bea
 J_{M}&=&\int\frac{Dy_{i}}{y_{1}y_{2}\bar{y}_{2}}~\int\frac{Dx_{i}}{x_{1}%
x_{2}\bar{x}_{2}}\,\frac{4\,x_{2}\,y_{2}}{D_{2}[\zeta]D_{2}[\bar{\zeta}%
]}\left\{  ~{e_{u}^{2}}\,\left[  (V^{\prime}+A^{\prime})(V+A)+4T^{\prime
}T\right]  (3,2,1)\right.  \nonumber\\
  & +& \left. e_{u}e_{d}\,\left[  (V^{\prime}+A^{\prime})(V+A)+4T^{\prime
}T\right]  (1,2,3)+e_{u}e_{d}\,2\left[  V^{\prime}V+A^{\prime}A\right]
(1,3,2)~\right\}  ,\label{JM}%
\eea
\bea
J_{3}&=&\int\frac{Dy_{i}}{y_{1}y_{2}\bar{y}_{2}\,}~\int\frac{Dx_{i}}{x_{1}%
x_{2}\bar{x}_{2}}\,\frac{2(x_{2}\,\bar{y}_{2}+\bar{x}_{2}\,y_{2})}{D_{2}%
[\zeta]D_{2}[\bar{\zeta}]}\left\{  {e_{u}^{2}}\,\left[  (V^{\prime}+A^{\prime
})(V+A)+4T^{\prime}T\right]  (3,2,1)\right.  \nonumber\\
&+&e_{u}e_{d}\,\left[  (V^{\prime}+A^{\prime})(V+A)+4T^{\prime}T\right]
(1,2,3)\left.  +e_{u}e_{d}\,2\left[  V^{\prime}V+A^{\prime}A\right]
(1,3,2)\right\} ,
\label{J3}%
\eea
where
\begin{equation}
Dy_{i}=dy_{1}dy_{2}dy_{3}\delta(1-y_{1}-y_{2}-y_{3}),
\end{equation}
quark charges $e_{u}=+2/3$, $e_{d}=-1/3$, $\alpha=e^{2}/(4\pi)$ and 
$\alpha_{s}(\mu^{2})$ is the strong coupling constant evaluated at scale
$\mu^{2}$. The  factor $D_{2}$ in the denominator  reads%
\begin{equation}
D_{2}[\zeta]=x_{2}\bar{\zeta}+y_{2}\zeta-x_{2}y_{2}+i\varepsilon
,~\ \ \zeta=\frac{1}{z}=\frac{s}{-t}>1.
\end{equation}
The nucleon DAs
\begin{equation}
V(x_{1},x_{2},x_{3})\equiv V(1,2,3)=\frac{1}{2}\left(  \varphi_{3}2,1,3)+\varphi_{3}(1,2,3)\right)  ,
\end{equation}%
\begin{equation}
A(1,2,3)=\frac{1}{2}\left(  \varphi_{3}(2,1,3)-\varphi_{3}(1,2,3)\right)  ,
\end{equation}%
\begin{equation}
T(1,2,3)=\frac{1}{2}\left[  V-A\right]  \left(  1,3,2\right)  +\frac{1}%
{2}\left[  V-A\right]  \left(  2,3,1\right) ,
\end{equation}
where the twist-3  DA  $\varphi_{3}$ is given in Eq.(\ref{phi3}). 
The products of the DAs in Eqs.(\ref{JM}) and (\ref{J3}) must be understood
as
\begin{equation}
\left[  V^{\prime}V+A^{\prime}A\right]  (1,3,2)=V^{\prime}%
(1,3,2)V(1,3,2)+A^{\prime}(1,3,2)A(1,3,2),
\end{equation}
with%
\begin{equation}
V^{\prime}(1,3,2)\equiv V(y_{1},y_{3},y_{2}),~\ V(1,3,2)\equiv V(x_{1}%
,x_{3},x_{2}).
\end{equation}

The expressions for the neutron amplitudes  can be obtained from
Eqs.(\ref{JM}) and (\ref{J3})   interchanging the  quark charges
$e_{u}\leftrightarrow e_{d}$. 

\section{The  amplitude $g_{1}$ within the EFT framework}
\label{appB}
The function $g_{1}$ describes  the contribution associated with the region
where one of the photons is soft and the other is hard.  We assume that in this case the virtual quark in the box diagram in Fig.\ref{soft-spectator} still has a sufficiently small virtuality and the interaction with the soft photon is described as a non-perturbative matrix element\begin{align}
& \left\langle p^{\prime},k^{\prime}\right\vert O_{+}^{\mu}~\bar{\psi
}(0)\gamma_{\mu}\psi(0)\left\vert k,p\right\rangle _{\text{EFT}}\nonumber\\
& =\bar{u}(k')\gamma_{\mu}u(k)~\bar{N}'\left[  \gamma^{\mu} f_{1}(Q^{2}) +  \gamma_{\sigma}\frac{\alpha}{\pi} \left(g_\bot^{\sigma\mu} g_{1}(\varepsilon,Q,\mu_{F}) +
\frac{{k}_{\bot}^{\sigma} k_{\bot}^{\mu}}{|k_{\bot}^{2}|}\ \frac{\alpha}{\pi}g_{3}(\varepsilon ,Q) \right)+\dots  \right]  N,
\label{def:g13}
\end{align}
where dots denote the possible chiral-odd structures which are not relevant now. 
Here the operator $O_{+}^{\mu}$ is constructed in SCET from the hard-collinear
fields, see the explicit expression in Eq.(\ref{Opl}). The operator
\ $\bar{\psi}(0)\gamma_{\mu}\psi(0)$ describes the lepton current. 
In the leading-order of electromagnetic coupling $\alpha=0$,  
Eq.(\ref{def:g13}) gives a matrix element that describes the soft-spectator
contribution within the SCET framework.   The soft photon exchange leads to the
two additional amplitudes $g_{1,3}$ in the {\it rhs} of Eq.(\ref{def:g13}). The
subscript EFT indicates that the matrix element is understood in the  effective
field theory, which describes interaction of the collinear hadrons with soft photon. 

In order to compute such a  matrix element one needs an effective field
framework. The hard subprocess in this picture  is described by the
interaction of the hard photon with the partons inside nucleon. Integration over hard modes
  gives the operator  $O_{+}^{\mu}$ describing  the vertices $\gamma^{\ast}p\rightarrow p$
or $\gamma^{\ast}p\rightarrow R$  where $R$ denotes an excited nucleon state.
We will assume that the partons after hard interactions  hadronise to a hadron, which
  interacts with the soft photon like a point-like particle. Corresponding vertex
$\gamma R\rightarrow p$ must be considered as a part of the  soft subprocess.
The soft photon can not resolve the structure of  hadrons and interacts with
the total  charge of a point-like particle. Therefore the corresponding vertex
can be approximated by the FF at  zero momentum transfer. The virtuality of
the soft photon in this framework  is restricted by the value factorisation scale
$\mu_{F}$,  which separates the  hard-hard and hard-soft photon-photon
configurations. 

This  effective field theory very close in spirit to the well-known hadronic
model  up to one point: computing the diagrams with the hadronic intermediate
state we assume that  virtualities of the particles in such effective theory
can not be large.   In order to guarantee
this requirement we neglect the small photon momentum components  in the
denominators of the propagators. This can be understood as an
expansion of the integrands in the diagrams of the hadronic model.  This is
somewhat similar  to the calculation of Mo and Tsai, except that we 
also to expand  the denominators. The obtained integrals have  UV-divergencies, which must be
absorbed into the evolution of the hard coefficient functions (which can be considered a a constants
 in the low-energy EFT).      

 In Ref.\cite{Kivel:2012vs} only the elastic contribution was calculated, which yields the equation (\ref{g1el}). This
term includes the QED IR-logarithm, which must cancel in the cross section. The
function $g_{1R}$ in Eq.(\ref{g1el}) includes the inelastic contributions.

\begin{figure}[ptb]%
\centering
\includegraphics[height=2.1287in,width=3.5293in]
{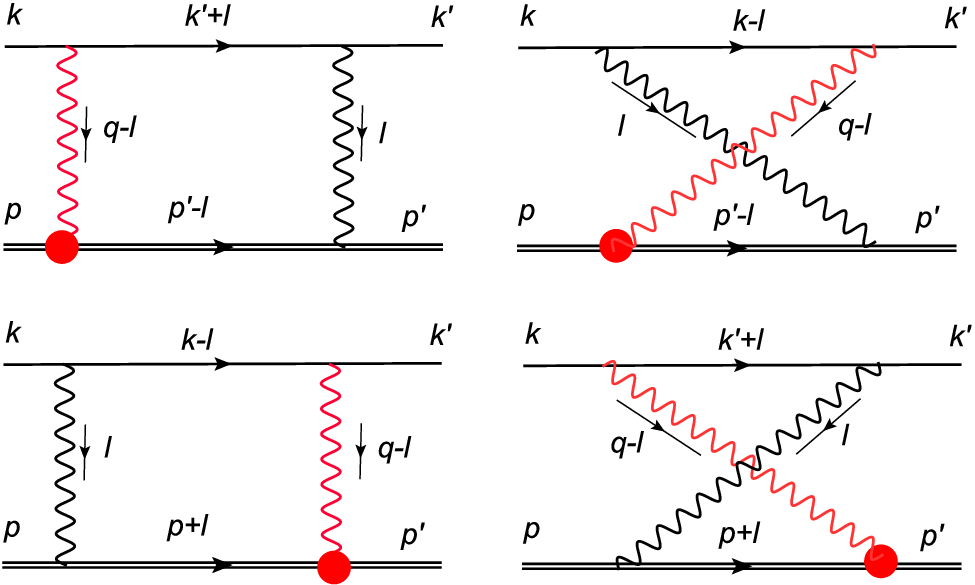}
\caption{The diagrams with the hard-soft photons. The hard lines are shown by red colour. }
\label{figresonans}
\end{figure}

Here we recalculate the function  $g_{1}$ and discuss the role of the resonance contributions. 
 We start our consideration from elastic term and reproduce the
result  in Eq.(\ref{g1el}).  We will perform our calculation using the
expansion of the hadronic diagrams. There are two diagrams which give four
contributions as shown in Fig.\ref{figresonans}. The hard photon is shown by
red colour, the red blob denotes the hard vertex.
Consider the elastic contribution.
\bea
\left\langle p^{\prime},k^{\prime}\right\vert O_{+}^{\mu}~\bar{\psi}
\gamma_{\mu}\psi\left\vert k,p\right\rangle _{\text{EFT,}el}  & = & 
G_F\int_{S}dl~
\bar{u} (k^{\prime})
\left[  \gamma_{\nu}\frac{(k^{\prime}+l)}{(k^{\prime}+l)^{2}}\gamma_{\mu}+\gamma_{\mu}\frac{(k-l)}{(k-l)^{2}}
\gamma_{\nu}\right]  
u(k)
\nonumber \\
& \times & \frac{1}{l^{2}-\lambda^{2}} \frac{1}{(q-l)^{2}}  \left(  M_{p1}^{\mu\nu}+M_{p2}^{\mu\nu} \right)  ,\label{<O>}
\eea
where, for simplicity, the factor $G_F$ absorbs  various  factors  from the  Feynman rules,  the photon mass $\lambda^{2}$ is used for the  IR-regularisation.  The functions $M_{pi}^{\mu\nu}$ describe the hadronic
subdiagrams of the first and second lines in Fig.\ref{figresonans},
respectively.  The symbol $\int_{S}$ $d^{4}l~$ in (\ref{<O>}) implies that 
we perfom integration over the region of the soft photon momentum
$l\lesssim\mu_{F}\sim\Lambda$. This means that one must expand the integrand
with respect to small $l$ in order to avoid the contributions  with  $l\gg\Lambda$.  
Such a  framework can be understood as a construction of the
effective field theory using the method known as expansion by regions \cite{Beneke:1997zp, Smirnov:2002pj}.
 The expansion of the integrand produces  UV-divergent integrals, which are
defined with the help of dimensional regularisation.  

Let us remind that we use in our description  the Breit frame, in which the
initial and final proton moves along $z$-axis and
\bea
p^{\prime}=p_{-}^{\prime}\frac{n}{2}+\frac{m}{p_{-}^{\prime}}\frac{\bar{n}}
{2},~\ p_{-}^{\prime}\sim Q,
\label{nvec}
\\
p=\frac{m}{p_{+}}\frac{n}{2}+p_{+}\frac{\bar{n}}{2},~\ p_{+}\sim Q.
\label{nbvec}
\eea
Here $n=(1,0,0,-1)$ and $\bar{n}=(1,0,0,1)$ are the auxiliary light-cone
vectors and $V_{\pm}=V_{0}\pm V_{3}$ as usually. 

The EFT framework implies that the nucleon spinors $\bar{N}(p^{\prime})$ and
$N(p)$ are reduced to the large components $\bar{N}^{\prime}$ and $N$,
respsectively, which  are defined as
\begin{equation}
\bar{N}^{\prime}=\bar{N}(p^{\prime})\frac{\nbs\ns}{4},~\ N=\frac
{\nbs\ns}{4}N(p).
\end{equation}

Performing  the expansion with respect to the small photon momentum  $l$ we are neglecting  all scalar products with small scales in the hard propagators, for  instance
\begin{align}
(q-l)^{2}  & \simeq q^{2},\label{(1stL)}
\end{align}
where we neglected  $l^{2}\sim \Lambda^{2}$ and $(ql)\sim\Lambda Q$ comparing to large (hard) $q^{2}$. We assume that the
hard interactions  are described  within the pQCD. 
In the hard-collinear lines we neglect by  soft terms 
\begin{align}
\frac{1}{(p+l)^{2}-m^{2}}  & \simeq\frac{1}{2(lp)},\label{(2ndL)}%
\end{align}
In the following  we suppose  that the long distance
physics  is associated with hard-collinear  $\Lambda Q\sim m$ and soft  $\Lambda^{2}$ scales.  This is valid only for a moderate values of $Q^2$, which guarantee that the hard-collinear scale is sufficiently small.  As a result  corresponding dynamics can be associated with  hadronic
degrees of freedom.  The virtualities of the corresponding particles in the
hadronic loop can not be higher than the hard-collinear scale. This restriction is provided by the expansion of the integrand
\begin{equation}
F(l,p,p^{\prime})\simeq F^{(\text{lo})}[l^{2},(lp),(lp^{\prime})]+\mathcal{O}(l/Q),
\end{equation}
where $F^{(\text{lo})}$   depends on the appropriate scalar products only.  After
that we can skip the restriction $S$ in the  loop integral and integrate over all phase space
 using   the  dimensional regularisation for treatment of the UV-divergencies  
\begin{equation}
\int_{S}d^{4}l~F(l,p,p^{\prime})\simeq\int d^{D}l~F^{(\text{lo}%
)}[l^{2},(lp),(lp^{\prime})],~ D=4-2\varepsilon.
\end{equation}
These   UV-divergencies must be understood in a sense of low energy effective field theory.  

 Performing expansion of the integrand in Eq.(\ref{<O>}) with respect to
 small photon momentum  $l$ we find
\bea
\left\langle p^{\prime},k^{\prime}\right\vert O_{+}^{\mu}\bar{\psi}%
\gamma_{\mu}\psi\left\vert k,p\right\rangle _{\text{EFT,}el}\simeq
GF\frac{\bar{u}^{\prime}\gamma_{\mu}u}{q^{2}}\int d^{D}l\left[  \frac{k^{\prime}_\nu%
}{(k^{\prime}l)}+\frac{k_{\nu}}{-(kl)}\right]  \frac{1}{l^{2}-\lambda^{2}%
}\left(  M_{p1}^{\mu\nu}+M_{p2}^{\mu\nu}\right),
\eea
with%
\begin{equation}
M_{p1}^{\mu\nu}=\bar{N}^{\prime}\left\{  \gamma^{\nu}G_{M}(0)-\frac
{p^{\prime\nu}+l^{\nu}/2}{m}F_{2}(0)\right\}  \frac{(p^{\prime}-l+m)}%
{(p^{\prime}-l)^{2}-m^{2}}\left\{  G_{M}(Q^{2})\gamma_{\bot}^{\mu}\right\}  N
\end{equation}%
\begin{align}
& \simeq G_{M}(Q^{2})~\bar{N}^{\prime}\gamma_{\bot}^{\mu}N~\left\{
G_{M}~\frac{p^{\prime\nu}}{-(p^{\prime}l)}-\frac{p^{\prime\nu}+l^{\nu}%
/2}{-(p^{\prime}l)}F_{2}\right\}  \\
& =G_{M}(Q^{2})~\bar{N}^{\prime}\gamma_{\bot}^{\mu}N~~\left[  F_{1}%
~\frac{p^{\prime\nu}}{-(p^{\prime}l)}-\frac{l^{\nu}}{-(p^{\prime}l)}\frac
{1}{2}F_{2}\right]  ~.
\end{align}

The similar consideration yields
\begin{equation}
M_{p2}^{\mu\nu}=G_{M}(Q^{2})~\bar{N}^{\prime}\gamma_{\bot}^{\mu}N~~\left[
F_{1}~\frac{p^{\nu}}{(pl)}+\frac{l^{\nu}}{(pl)}\frac{1}{2}F_{2}\right]  .
\end{equation}
Taking into account the gauge invariance
\begin{equation}
\left[  \frac{k^{\prime}_\nu}{(k^{\prime}l)}+\frac{k_{\nu}}{-(kl)}\right]
l^{\nu}=0,\label{cont}
\end{equation}
and normalisation $F_{1}(0)=1$ we obtain
\begin{equation}
M_{1h}^{\mu\nu}+M_{2h}^{\mu\nu}=G_{M}(Q^{2})~\bar{N}^{\prime}\gamma_{\bot
}^{\mu}N~~\left[  \frac{p^{\nu}}{(pl)}+\frac{p^{\prime\nu}}{-(p^{\prime}%
l)}\right]  .
\end{equation}
Therefore  we  find
\bea
 \left\langle p^{\prime},k^{\prime}\right\vert O_{+}^{\mu}~\bar{\psi}%
\gamma_{\mu}\psi\left\vert k,p\right\rangle _{\text{EFT,}el} &\simeq&
GF~\frac{\bar{u}^{\prime}\gamma_{\mu}u}{q^{2}}\bar{N}^{\prime}\gamma_{\bot}^{\mu}N~G_{M}(Q^{2})
\nonumber \\
& \times&\int\frac{d^{D}l}{l^{2}-\lambda^{2}}~\left[  \frac{k^{\prime}_\nu
}{(k^{\prime}l)}+\frac{k_{\nu}}{-(kl)}\right]  \left[  \frac{p^{\nu}}%
{(pl)}+\frac{p^{\prime\nu}}{-(p^{\prime}l)}\right]  .\label{soft-el}%
\eea
Calculation of the integrals in (\ref{soft-el}) gives expression in
Eq.(\ref{g1el}).

Consider now inelastic contribution with the resonance $J^{P}=$ $\frac{1}{2}^{-}.$
The transition vertices are described as \cite{Devenish:1975jd}
\begin{align}
V_{\gamma p\rightarrow R}^{\mu} =G_{1}(Q^{2})\left(  q^{\sigma}q^{\mu
}-q^{2}g^{\mu\sigma}\right)  \gamma_{\sigma} +G_{2}(Q^{2})\left[  q^{\sigma}\bar{P}^{\nu}-(\bar{P}q)g^{\mu\sigma}\right]
\gamma_{\sigma},
\end{align}
where $q$ is incoming photon momentum and $\bar P=\frac{1}{2}(p+p_{R})$.
This gives%
\bea
\left.  \left\langle p^{\prime},k^{\prime}\right\vert O_{+}^{\mu}~\bar{\psi
}\gamma_{\mu}\psi\left\vert k,p\right\rangle _{\text{EFT}}\right\vert
_{\frac{1}{2}^{-}}\simeq GF~\frac{\bar{u}^{\prime}\gamma^{\mu}u}{q^{2}%
} \int dl~\left[  \frac{k^{\prime}_\nu
}{(k^{\prime}l)}+\frac{k_{\nu}}{-(kl)}\right]  \frac{1}{l^{2}}\left(
M_{R1}^{\mu\nu}+M_{R2}^{\mu\nu}\right) ,
\label{me12}%
\eea
with%
\begin{align}
M_{R1}^{\mu\nu}  & \simeq G_{1}(0)\left(  l^{\sigma}l^{\nu}-l^{2}g^{\nu\sigma
}\right)  ~\bar{N}^{\prime}\gamma_{\sigma}\frac{(\Dsl{p}^{\prime}-\Dsl{l}+m_{R}%
)}{(p^{\prime}-l)^{2}-m_{R}^{2}}\left\{  Q^{2}G_{1}(Q^{2})\gamma_{\bot}^{\mu
}\right\}  N\nonumber\\
& +G_{2}(0)\left\{  l^{\sigma}\bar{P}^{\nu}-(\bar{P}l)g^{\nu\sigma}\right\}  \bar{N}^{\prime
}\gamma_{\sigma}\frac{(\Dsl{p}^{\prime}-\Dsl{l}+m_{R})}{(p^{\prime}-l)^{2}-m_{R}^{2}%
}\left\{  -q^{2}G_{1}(Q^{2})\gamma_{\bot}^{\mu}\right\}  N,\label{M1R}%
\end{align}
where%
\begin{equation}
\Delta^{2}=m_{R}^{2}-m^{2},~ \bar{P}=p^{\prime}-l/2.
\end{equation}
Similar expression also holds for $M_{R2}^{\mu\nu}$.   Consider the expansion
with respect to small $l$. The first line in {\it rhs} of (\ref{M1R}) gives
\begin{align}
&G_{1}(0) \left(  l^{\sigma}l^{\nu}-l^{2}g^{\nu\sigma}\right)
~\bar{N}^{\prime}\gamma_{\sigma}\frac{(\Dsl{p}^{\prime}-\Dsl{l}+m_{R})}{(p^{\prime}%
-l)^{2}-m_{R}^{2}}\gamma_{\bot}^{\mu}N\\
& \simeq~\bar{N}^{\prime}\gamma_{\bot}^{\mu}N~G_{1}(0)\frac{(lp^{\prime})l^{\nu
}-l^{2}p^{\prime\nu}}{-2(p^{\prime}l)-\Delta^{2}}.
\end{align}
In the expansion of the hard-collinear propagator we keep the mass difference $\Delta^2$ because numerically this 
difference is of order of the hard-collinear scale.  
The contribution with $l^{\nu}$ vanishes due to contraction as in  Eq.(\ref{cont}).
 The term with $l^{2}$ gives the scaleless integral
\begin{equation}
\int dl~\left[  \frac{k^{\prime}_\nu}{(k^{\prime}l)}+\frac{k_{\nu}}%
{-(kl)}\right]  \frac{p^{\prime\nu}}{-2(p^{\prime}l)-\Delta^{2}}=0.
\label{intM1}%
\end{equation}
These integrals are scaleless  with respect to transverse
components $d^{D-2}l_{\top}$\footnote{We reserve the symbol $\bot$ for the transverse component with respect to  $n$ and $\bar{n}$ vectors in (\ref{nvec}) and (\ref{nbvec}). For another light-cone basis vectors we use symbol $\top$.  }, which can be easily obtained performing the  light-cone
expansion  with respect to external momenta in each case. Therefore  integral
(\ref{intM1}) vanishes.  

  The second line in Eq.(\ref{M1R}) gives%
\bea
&&\left\{  l^{\sigma}\bar{P}^{\nu}-(\bar{P}l)g^{\nu\sigma}\right\}  \bar
{N}^{\prime}\gamma_{\sigma}\frac{(\Dsl{p}^{\prime}-\Dsl{l}+m_{R})}{(p^{\prime}
-l)^{2}-m_{R}^{2}}\gamma_{\bot}^{\mu}N
\\ &&
\simeq\frac{\left(  l^{\sigma}\bar{P}^{\nu}-(\bar{P}l)g^{\nu\sigma}\right)  p^{\prime
}_\sigma}{-2(p^{\prime}l)-\Delta^{2}}~\bar{N}^{\prime}\gamma_{\bot}^{\mu
}N+\frac{\left(  l^{\sigma}\bar{P}^{\nu}-(\bar{P}l)g^{\nu\sigma}\right)  }{-2(p^{\prime}l)-\Delta^{2}}\bar{N}^{\prime}\gamma_{\sigma}\not l\gamma_{\bot}^{\mu
}N.\label{M12R}
\eea
The first term in (\ref{M12R}) gives
\begin{equation}
\left(  l^{\sigma}\bar{P}^{\nu}-(\bar{P}l)g^{\nu\sigma}\right)  2p^{\prime}_\sigma=\frac
{1}{2}\left[  l^{2}p^{\prime\nu}-(lp^{\prime})l^{\nu}\right]  ,
\end{equation}
which also yields the scaleless integrals as in Eq.(\ref{intM1}). The second
term in (\ref{M12R}) gives
\begin{align}
& \frac{\left(  l^{\sigma}\bar{P}^{\nu}-(\bar{P}l)g^{\nu\sigma}\right)  l^{\lambda}%
}{-2(p^{\prime}l)-\Delta^{2}}\bar{N}^{\prime}\gamma_{\sigma}\gamma_{\lambda
}\gamma_{\bot}^{\mu}N\\
& =\frac{\left(  l^{2}\bar{P}^{\nu}-(\bar{P}l)l^{\nu}\right)  }{-2(p^{\prime}l)-\Delta
^{2}}\bar{N}^{\prime}\gamma_{\bot}^{\mu}N+\frac{-(\bar{P}l)g^{\nu\sigma}l^{\lambda}%
}{-2(p^{\prime}l)-\Delta^{2}}\frac{1}{2}\bar{N}^{\prime}\left[  \gamma_{\sigma},\gamma_{\lambda}\right]  \gamma_{\bot}^{\mu}N
\\
& \simeq\frac{-(p^{\prime}l)g^{\nu\sigma}l^{\lambda}}{-2(p^{\prime}l)-\Delta
^{2}}\frac{1}{2}\bar{N}^{\prime}\left[  \gamma_{\sigma},\gamma_{\lambda
}\right]  \gamma_{\bot}^{\mu}N.
\end{align}
where we again neglected the terms with $l^{2}$ and $l^{\nu}$. The remaining
term gives
\bea
 &&\frac{\bar{u}^{\prime}\gamma_{\mu}u}{q^{2}}\int dl~\left[  \frac
{k^{\prime}_\nu}{(k^{\prime}l)}+\frac{k_{\nu}}{-(kl)}\right]  \frac{1}{l^{2}%
}M_{N1}^{\mu\nu}%
\\
&&\sim~\frac{\bar{u}^{\prime}\gamma^{\mu}u}{-Q^{2}}\frac{1}{2}\bar{N}^{\prime
}\gamma_{\bot}^{\mu}\left[  \gamma_{\lambda},\gamma_{\nu}\right]  N\int
dl~\left[  \frac{k^{\prime\nu}}{(k^{\prime}l)}+\frac{k^{\nu}}{-(kl)}\right]
\frac{1}{l^{2}}\frac{-(lp^{\prime})l^{\lambda}}{-2\left(  p^{\prime}l\right)
-\Delta^{2}}%
\\ &&
\sim\frac{\bar{u}^{\prime}\gamma^{\mu}u}{-Q^{2}}\frac{1}{2}\bar{N}^{\prime
}\gamma_{\bot}^{\mu}\left[  \gamma_{\lambda},\gamma_{\nu}\right]
N\frac{\Delta^{2}}{2}\int dl~\left[  \frac{k^{\prime\nu}}{(k^{\prime}l)}%
+\frac{k^{\nu}}{-(kl)}\right]  \frac{1}{l^{2}}\frac{l^{\lambda}}{-2\left(
p^{\prime}l\right)  -\Delta^{2}},
\eea
where we neglected the scaleless integral%
\begin{equation}
\int dl~\left[  \frac{k^{\prime\nu}}{(k^{\prime}l)}+\frac{k^{\nu}}%
{-(kl)}\right]  \frac{1}{l^{2}}=0.
\end{equation}
In order to analyse the remnant terms  we perform the light-cone expansions of the
$l^{\lambda}$ assuming in the numerator $p^{\prime2}\simeq0,\ k^2\simeq0,\ k'^2\simeq0$:%
\begin{align}
l^{\lambda}  & =p^{\prime\lambda}\frac{(kl)}{(p^{\prime}k)}+k^{\lambda}%
\frac{(lp^{\prime})}{(p^{\prime}k)}+l_{\top}^{\lambda^{\prime}},\\
~l^{\lambda}  & =p^{\prime\lambda}\frac{(k^{\prime}l)}{(p^{\prime}k^{\prime}%
)}+k^{\prime\lambda}\frac{(lp^{\prime})}{(p^{\prime}k^{\prime})}+l_{\top
}^{\lambda^{\prime}}\ .
\end{align}
The  terms with transverse component $l_{\top}$ vanish due to rotation
invariance. The terms with $k^{\sigma}k^{\lambda}$ and $k^{\prime\sigma
}k^{\prime\lambda}$ vanish because of contraction with $\bar{N}^{\prime}%
\gamma_{\bot}^{\mu}\left[  \gamma^{\lambda},\gamma^{\nu}\right]  N$. \ Hence
we only get the combinations%
\begin{align}
p^{\prime\lambda}\left[  \frac{k^{\prime\nu}}{(p^{\prime}k^{\prime})}%
-\frac{k^{\nu}}{(p^{\prime}k)}\right]  \int d^{D}l~\frac{1}{l^{2}}\frac
{1}{-2\left(  p^{\prime}l\right)  -\Delta^{2}}.
\end{align}
The contraction of the Lorentz indices yields%
\bea
\bar{N}^{\prime}\gamma_{\bot}^{\mu}\left[  \gamma_{\lambda},\gamma_{\nu
}\right]  N~p^{\prime\lambda}\left[  \frac{k^{\prime\nu}}{(p^{\prime}%
k^{\prime})}-\frac{k^{\nu}}{(p^{\prime}k)}\right]  =-\bar{N}^{\prime}%
\gamma_{\bot}^{\mu}\gamma_{\nu}\Dsl{p}^{\prime}N\left[  \frac{k^{\prime\nu}%
}{(p^{\prime}k^{\prime})}-\frac{k^{\nu}}{(p^{\prime}k)}\right]
\eea
\begin{equation}
=-\bar{N}^{\prime}\gamma_{\bot}^{\mu}N~2p^{\prime}_\nu\left[  \frac
{k^{\prime\nu}}{(p^{\prime}k^{\prime})}-\frac{k^{\nu}}{(p^{\prime}k)}\right]
=0.
\end{equation}
Therefore %
\begin{equation}
GF~\frac{\bar{u}^{\prime}\gamma_{\mu}u}{q^{2}}\int d^{D}l~\left[
\frac{k^{\prime}_\nu}{(k^{\prime}l)}+\frac{k_{\nu}}{-(kl)}\right]  \frac
{1}{l^{2}}M_{R1}^{\mu\nu}=0.
\end{equation}
The similar analysis for the second contribution with $M_{R2}^{\mu\nu}$ in
Eq.(\ref{me12})  gives%
\begin{equation}
GF~\frac{\bar{u}^{\prime}\gamma_{\mu}u}{q^{2}}\int d^{D}l~\left[
\frac{k^{\prime}_\nu}{(k^{\prime}l)}+\frac{k_{\nu}}{-(kl)}\right]  \frac
{1}{l^{2}}M_{R2}^{\mu\nu}=0.
\end{equation}
Therefore
\begin{equation}
\left.  \left\langle p^{\prime},k^{\prime}\right\vert O_{+}^{\mu}~\bar{\psi
}\gamma^{\mu}\psi\left\vert k,p\right\rangle _{\text{EFT}}\right\vert
_{\frac{1}{2}^{-}}\simeq 0.
\end{equation}

Let us also consider  the contribution of the $\Delta$-resonance.
 For simplicity,  we restrict our consideration to the leading magnetic dipole
transition. Corresponding vertex can be written as \cite{Jones:1972ky}
\begin{align}
V_{\gamma p\rightarrow\Delta}^{\alpha\mu}  & =\sqrt{\frac{3}{2}}\frac{\left(
1+m_{\Delta}/m\right)  }{\left(  1+m_{\Delta}/m\right)  ^{2}+4\tau^{2}}%
\frac{G^{\ast}(Q^{2})}{m^{2}}\varepsilon^{\alpha\mu\rho\sigma}(p_{\Delta
})_{\rho}\, p_{\sigma}\\
& =V_{\gamma p\rightarrow\Delta}~(Q^{2})~\varepsilon^{\alpha\mu\rho\sigma
}(p_{\Delta})_{\rho}\, p_{\sigma}.
\end{align}
where  the incoming photon momentum defined as $q=p_{\Delta}-p$.

The matrix element reads, see {\it e.g.}  \cite{Tomalak:2014sva} %
\bea
\left.  \left\langle p^{\prime},k^{\prime}\right\vert O_{+}^{\mu}~\bar{\psi
}\gamma_{\mu}\psi\left\vert k,p\right\rangle _{\text{EFT}}\right\vert
_{\frac{3}{2}^{+}}& \simeq &  G_F~\frac{\bar{u}^{\prime}\gamma_{\mu}u}{q^{2}%
}\nonumber\\
&\times& \int d^{D}l~\left[  \frac{k^{\prime}_\nu
}{(k^{\prime}l)}+\frac{k_{\nu}}{-(kl)}\right]  \frac{1}{l^{2}}\left(
M_{\Delta1}^{\mu\nu}+M_{\Delta2}^{\mu\nu}\right)  .\label{<O>dlt}%
\eea
The first  contribution reads%
\bea
M_{\Delta1}^{\mu\nu}&=&\bar{N}^{\prime}\left(  \gamma_{0}V_{p\Delta}^{\beta\nu
}(l)\gamma_{0}\right)  ^{\dag}\frac{\Dsl{p}^{\prime}+\Dsl{l}+m_{\Delta}}{(p^{\prime
}+l)^{2}-m_{\Delta}^{2}}\left(  -g_{\alpha\beta}+\frac{1}{3}\gamma_{\beta
}\gamma_{\alpha}\right)  V_{p\Delta}^{\alpha\mu}(q)N
\\
&=&V_{p\Delta}(0)V_{p\Delta}(Q^{2})
~\frac{\varepsilon^{\alpha\mu p^{\prime}p}\varepsilon^{\beta\nu p^{\prime}l}}{-2(p^{\prime}l)-\Delta^{2}}\left(-l^{\sigma}\right) 
 \bar{N}^{^{\prime}}\gamma_{\sigma}\left(  -g_{\alpha\beta}+\frac{1}{3}\gamma_{\beta}\gamma_{\alpha}\right)  N.
\eea
where we neglected contributions with the chiral-odd structures~$\bar
{N}^{\prime}N$, which we do not need and  $\Delta^{2}=m_{\Delta}^{2}-m^{2}$.
 This expression gives the contribution, which is proportional to the
following  product%
\begin{equation}
\varepsilon^{\alpha\mu p^{\prime}p}\varepsilon^{\beta\nu p^{\prime}\lambda}
\bar{N}^{^{\prime}}\gamma^{\sigma}\left(  -g_{\alpha\beta}+\frac
{1}{3}\gamma_{\beta}\gamma_{\alpha}\right)  N~J_{\nu\lambda\sigma
},\label{epsJ123}%
\end{equation}
with%
\begin{equation}
~J_{\nu\lambda\sigma}=\int dl~\frac{1}{\left[  l^{2}\right]  }\left[
\frac{k^{\prime}_\nu}{(k^{\prime}l)}+\frac{k_{\nu}}{-(kl)}\right]
\frac{l_{\lambda}l_{\sigma}}{-2(p^{\prime}l)-\Delta^{2}}.
\end{equation}
We again use the appropriate light-cone expansions  for each integral %
\bea
&&l_{\sigma}\simeq p^{\prime}_\sigma\frac{(lk)}{(p^{\prime}k)}+k_{\sigma}%
\frac{(lp^{\prime})}{(kp^{\prime})}+l_{\top}^{\prime},
\\
&& l_{\sigma}\simeq p^{\prime}_\sigma\frac{(lk^{\prime})}{(p^{\prime}k^{\prime}%
)}+k^{\prime}_\sigma\frac{(lp^{\prime})}{(k^{\prime}p^{\prime})}+l_{\top}.
\eea
Then the contributions with $p^{\prime}_\sigma$ vanish because of contraction
with  $\bar{N}^{^{\prime}}\gamma^{\sigma}\left(  \ldots\right)N$ Therefore we obtain%
\bea
J_{\nu\lambda\sigma}&\simeq& \int dl~\frac{1}{\left[  l^{2}\right]  }\left[
\frac{k^{\prime}_\nu k^{\prime}_\sigma}{(k^{\prime}l)}+\frac{k_{\nu}k_{\sigma}%
}{-(kl)}\right]  \frac{l_{\lambda}(lp^{\prime})}{-2(p^{\prime}l)-\Delta^{2}}%
\\
&+&\int dl~\frac{1}{\left[  l^{2}\right]  }\left[  \frac{k^{\prime\nu}l_{\top\sigma}^{\prime}}{(k^{\prime}l)}+\frac{k_{\nu}l_{\top\sigma}}%
{-(kl)}\right]  \frac{l_{\lambda}}{-2(p^{\prime}l)-\Delta^{2}}.
\eea
Performing the same expansion for $l_{\lambda}$ and taking into account
rotation invariance  and also using that the terms with $k_{\nu}k_{\lambda}$ and
$k^\prime_{\nu}k^\prime_{\lambda}$  or with $p^{\prime}_\lambda$ vanish due to
contraction with $\varepsilon^{\beta\nu p^{\prime}\lambda}$ in
(\ref{epsJ123}) we find%
\bea
J_{\nu\lambda\sigma}\sim\int dl~\frac{1}{\left[  l^{2}\right]  }\left[
k^{\prime}_\nu (g_{\top})_{\sigma\lambda}^{\prime}\frac{l_{\top}^{\prime2}%
}{(k^{\prime}l)}+k_{\nu}(g_{\top})_{\sigma\lambda}\frac{l_{\top}^{2}}%
{-(kl)}\right]  \frac{1}{-2(p^{\prime}l)-\Delta^{2}}.
\eea
The UV-divergent integrals with $l_{\top}^{\prime2}$ and $l_{\top}^{2}$ can be
understood using that
\begin{equation}
l_{\top}^{2}=l^{2}-(p^{\prime}l)(kl)\frac{1}{(p^{\prime}k)},~\ l_{\top
}^{\prime2}=l^{2}-(p^{\prime}l)(k^{\prime}l)\frac{1}{(p^{\prime}k^{\prime})}.
\end{equation}
The terms in rhs of the both expressions give the scaleless integrals   only
and therefore also vanish. Hence we obtain that the contribution with
$M_{\Delta1}^{\mu\nu}$ in Eq.(\ref{<O>dlt}) gives trivial contribution. The
same  result can be obtained  in the same way  for $M_{\Delta2}^{\mu\nu}$ in
 Eq.(\ref{<O>dlt}). Therefore
\begin{equation}
\left.  \left\langle p^{\prime},k^{\prime}\right\vert O_{+}^{\mu}~\bar{\psi
}\gamma_{\mu}\psi\left\vert k,p\right\rangle _{\text{EFT}}\right\vert
_{\frac{3}{2}^{+}}\simeq 0.
\end{equation}
The similar consideration can also be used  for the other resonances with
$J^{P}=\frac{1}{2}^{+},\frac{3}{2}^{-}$.  Therefore we conclude that we do
not obtaind  the relevant contributions from the excited states and
\ therefore
\begin{equation}
\left.  g_{1}(\varepsilon,Q^{2})\right\vert _{R}=0,~\ g_{3}(\varepsilon
,Q^{2})=0.
\end{equation}

\section{Estimates of the neutron FF $\mathcal{F}_{1n}$}
\label{appC}
In this section we evaluate the FF $\mathcal{F}_{1n}$ that enters into the description of the TPE amplitudes for the neutron target.   We use  isotopic symmetry and known  electromagnetic FFs for the proton and neutron and proton  FF $\mathcal{F}_{1}$  defined from the WACS data.  

We start  from the definition of these FFs in the EFT approach.  For the
hard-collinear fields we use the same notations as in Ref. \cite{Kivel:2012vs}. The auxiliary
light-cone vectors $n$ and $\bar{n}$ are associated with the momenta
$p^{\prime}$ and $p$, respectively,  as defined in Eqs.(\ref{nvec}) and  (\ref{nbvec}).  Remind that  we assume the Breit frame as in
Ref. \cite{Kivel:2012vs}. We also use the convenient notation for the gauge invariant
combinations defined as 
\begin{equation}
\chi_{n}(\lambda\bar{n})\equiv~W_{n}(\lambda\bar{n})\xi_{n}(\lambda\bar
{n}),\label{qjet}%
\end{equation}
where the hard-collinear gluon Wilson line (WL) reads~:
\begin{equation}
W_{n}(z)=\text{P}\exp\left\{  ig\int_{-\infty}^{0}ds~\bar{n}\cdot
A^{(n)}(z+s\bar{n})\right\}  .
\end{equation}

Assuming that the nucleon electromagnetic FFs are dominated by the
soft-spectator contribution in the region where the hard-collinear scale is
not large $\Lambda Q<1$GeV$^{2}$, it is suggested that the Dirac FFs can be
approximated by the corresponding matrix elements
\begin{equation}
F_{1}^{p}(Q^{2})\simeq f_{1p}(Q^{2}),~\ F_{1}^{n}(Q^{2})\simeq f_{1n}%
(Q^{2}),\label{F1p}%
\end{equation}
where the SCET FFs are defined as%
\begin{equation}
\left\langle P(p^{\prime})\right\vert O_{+}^{\mu}~\left\vert P(p)\right\rangle
_{\text{{\footnotesize SCET}}}=\bar{N}'\gamma_{\bot}^{\mu}N~f_{1p}(Q^{2}),\label{def:f1}%
\end{equation}
and the SCET operator reads%
\begin{equation}
O_{+}^{\mu}=\sum_{q=u,d}e_{q}\left\{  \bar{\chi}_{n}(0)\gamma_{\bot}^{\mu
}~\chi_{\bar{n}}(0)+\bar{\chi}_{\bar{n}}(0)\gamma_{\bot}^{\mu}\chi
_{n}(0)\right\}  ,\label{Opl}%
\end{equation}
where $e_{u}=2/3$ and $e_{d}=-1/3$. The similar definition also holds for the
neutron state. The matrix element (\ref{def:f1}) describes the configuration with
one active quark or antiquark and soft spectators, see {\it e.g.} Ref.\cite{Kivel:2010ns}.  The
first term $\bar{\chi}_{n}\gamma_{\bot}^{\mu}~\chi_{\bar{n}}$ can be
associated with the active quark, the second one$~\bar{\chi}_{\bar{n}}%
\gamma_{\bot}^{\mu}\chi_{n}$ with active antiquark. Therefore we can introduce the following decomposition 
\begin{equation}
~f_{1p}=e_{u}(f_{1p}^{u}-\bar{f}_{1p}^{u})+e_{d}(f_{1p}^{d}-\bar{f}_{1p}%
^{d}),\label{f1p}%
\end{equation}
where$~f_{1}^{u,d}$ and $\bar{f}_{1}^{u,d}$ describe the contributions of the
quark and antiquarks, respectively. \ \ Similarly
\begin{equation}
~f_{1n}=e_{u}(f_{1n}^{u}-\bar{f}_{1n}^{u})+e_{d}(f_{1n}^{d}-\bar{f}_{1n}%
^{d}).\label{f1n}%
\end{equation}
The isotopic symmtry implies that%
\begin{equation}
f_{1p}^{u}=f_{1n}^{d}\equiv f_{1}^{u},~\ f_{1p}^{d}=f_{1n}^{u}\equiv
\ f_{1}^{d},
\end{equation}%
\begin{equation}
\bar{f}_{1p}^{u}=\bar{f}_{1n}^{d}\equiv\bar{f}_{1}^{u},~\ \bar{f}_{1p}%
^{d}=\bar{f}_{1n}^{u}\equiv\ \bar{f}_{1}^{d}.
\end{equation}

Let us briefly remind, that these soft-overlap  FFs arise in SCET with the  coefficient function
 of order one in $\alpha_{s}$ . These FFs still depend on the large momentum transfer $Q^{2}$,
but this behaviour is defined by the hard-collinear interactions associated with the
typical scale of order $\Lambda Q$. Because this scale is assumed to be
relatively small we can not develop factorisation for the SCET FFs and
consider these quantities as non-pertubative.  

In the asymptotic limit  $Q\rightarrow\infty$   behaviour of the quark and
antiquark FFs is different. The antiquarks can not appear in the leading power
contribution because the minimal Fock state is described by three
quarks only. As a result the antiquark FFs are 
suppressed by additional powers of  $1/Q^{2}$. However,  in the region
of moderate values of the hard scale, where the hard-collinear scale is still relatively small,  we
suppose that the antiquark FFs are still not quite small. In what follow
we\emph{ assume} that antiquark form factors  for  $u$- and $d$-quarks are
approximately the same
\begin{equation}
\bar{f}_{1}^{u}\simeq\bar{f}_{1}^{d}\equiv\bar{f}_{1}.
\end{equation}
Then using (\ref{F1p}) and the flavor decompositions (\ref{f1p}) and
(\ref{f1n}) we get
\begin{align}
F_{1p}  & \simeq e_{u}(f_{1}^{u}-\bar{f}_{1})+e_{d}(f_{1}^{d}-\bar{f}%
_{1})=e_{u}f_{1}^{u}+e_{d}f_{1}^{d}-(e_{u}+e_{d})\bar{f}_{1},\\
F_{1n}  & \simeq e_{u}(f_{1}^{d}-\bar{f}_{1})+e_{d}(f_{1}^{u}-\bar{f}%
_{1})=e_{u}f_{1}^{d}+e_{d}f_{1}^{u}-(e_{u}+e_{d})\bar{f}_{1}.
\end{align}
 This  can also be rewritten as
\begin{equation}
2F_{1p}+F_{1n}\equiv F_{1}^{u}=f_{1}^{u}-\bar{f}_{1},\label{F1u}%
\end{equation}%
\begin{equation}
F_{1p}+2F_{1n}\equiv F_{1}^{d}=f_{1}^{d}-\bar{f}_{1},\label{F1d}%
\end{equation}
where we substitute the explicit values of the quark charges. The FFs
$F_{1}^{u}$ and $F_{1}^{d}$ are often referred in the literature as
electromagnetic FF's of $u$- and $d$-quark, respectively. Using the data for the
proton and neutron FFs these  $F_{1}^{q}$ can be easily obtained,  see {\it e.g.} Ref.\cite{Cates:2011pz}.

The  WACS form factors  are defined as \cite{Kivel:2012vs}
\begin{equation}
\left\langle P(p^{\prime})\right\vert O_{-}^{\mu}~\left\vert P(p)\right\rangle
_{\text{{\footnotesize SCET}}}=\bar{N}'\gamma_{\bot}^{\mu}N~\mathcal{F}_{1p}(Q^{2}),
\end{equation}
where the SCET operator reads%
\begin{equation}
O_{-}^{\mu}=\sum_{q=u,d}e_{q}^{2}\left\{  \bar{\chi}_{n}(0)\gamma_{\bot}^{\mu
}~\chi_{\bar{n}}(0)-\bar{\chi}_{\bar{n}}(0)\gamma_{\bot}^{\mu}\chi_{n}(0)\right\}  .
\end{equation}
The same definition is implied for the neutron FF $\mathcal{F}_{1n}(Q^{2})$,
which we need to calculate.  Now the antiquark components  enter with the sign minus
because the corresponding operator is $C$-even.  Using the quark and antiquark FFs we
obtain%
\begin{equation}
\mathcal{F}_{1p}\equiv\mathcal{F}_{1}=e_{u}^{2}~f_{1}^{u}+e_{d}^{2}f_{1}%
^{d}+\left(  e_{u}^{2}+e_{d}^{2}\right)  \bar{f}_{1}.
\end{equation}%
\begin{equation}
\mathcal{F}_{1n}=e_{u}^{2}~f_{1}^{d}+e_{d}^{2}f_{1}^{u}+\left(  e_{u}%
^{2}+e_{d}^{2}\right)  \bar{f}_{1}.
\end{equation}
The data for the proton WACS allows one to get information about  FF
$\mathcal{F}_{1}$  \cite{Kivel:2012vs, Kivel:2015vwa}, which can be parametrised as in Eq.(\ref{Fwacs}).  Then
using (\ref{F1u}) and (\ref{F1d})  one finds%
\begin{equation}
f_{1}^{u}=\frac{9}{10}\mathcal{F}_{1}+\frac{6}{10}F_{1}^{u}-\frac{1}{10}%
F_{1}^{d},
\end{equation}%
\begin{equation}
f_{1}^{d}=\frac{9}{10}\mathcal{F}_{1}-\frac{4}{10}F_{1}^{u}+\frac{9}{10}%
F_{1}^{d},
\end{equation}%
\begin{equation}
~\bar{f}_{1}=\frac{9}{10}\mathcal{F}_{1}-\frac{4}{10}F_{1}^{u}-\frac{1}%
{10}F_{1}^{d}.
\end{equation}
Using these relations one finds%
\begin{equation}
\mathcal{F}_{1n}=\mathcal{F}_{1}+\frac{1}{3}\left(  F_{1}^{d}-F_{1}%
^{u}\right)  =\mathcal{F}_{1}+\frac{1}{3}\left(  F_{1n}-F_{1p}\right)  .
\end{equation}
The flavour separation has been considered in Ref.\cite{Cates:2011pz}. The existing data allow
one only to consider two points  in the region  $Q^{2}$ $\geq2.5$ GeV$^{2}$.  Using these
information  we find the results for quark, antiquark and neutron FF  $\mathcal{F}%
_{1n}$, which are presented in Table~\ref{tab1}.

\begin{table}[th]
\centering
\begin{tabular}{|c|c |c |c |c |c |c|c| }
\hline
$Q^{2}$, GeV$^{2}$ & $F_{1}^{u}$ & $F_{1}^{d}$ & $\mathcal{F}_{1}$ &
$f_{1}^{u}$ & $f_{1}^{d}$ & $\bar{f}_{1}$ & $\mathcal{F}_{1n}$\\ \hline
$2.5$ & $0.14\,$ & $0.028$ & $0.29$ & $0.34$ & $0.23$ & $0.20$ & $0.25~$\\ \hline
$3.4$ & $0.085$ & $0.013$ & $0.15$ & $0.19$ & $0.11$ & $0.10$ & $0.12$\\
\hline
\end{tabular}
\caption{ Results for quark and antiquark FFs, as well as for the neutron FF $\mathcal{F}_{1n}$. }  
\label{tab1}
\end{table}

 From this table it is seen that the antiquark FF is quite large and comparable
to the quark FFs $f_{1}^{q}$.  
This result is a consequence of the rather small e.m. $d$-quark FF: $F_{1}^{d}\ll F_{1}^{u}$. 
In the soft-spectator picture, this fact is naturally interpreted as a cancellation between the FF of the quark and antiquark in the region of moderate values of $Q^{2}$.  

Remind,  the antiquark FF is formally power suppressed as it follows from the naive power counting but  for the soft-spectator scattering the power behaviour is associated  with the hard-collinear scale, which is assumed to be relatively small . The hard coefficient functions for  quark and antiquark  FFs are of order one
with respect to  $\alpha_{s}$. They have NLO corrections of order $\alpha_{s}$,  which have been 
neglected  for simplicity. The hard-spectator contributions have the hard coefficient functions of order
$\alpha_{s}^{2}$ and  have also been  neglected.

\end{document}